\documentclass[10pt,a4paper]{article}
\pdfoutput=1
 \usepackage{jcappub}

\allowdisplaybreaks

\usepackage{ifthen}

\usepackage{amsmath} 
\usepackage{amssymb}
\usepackage{amsthm}
\usepackage{amsfonts}
\usepackage{array}
\usepackage{multirow}
\usepackage{bm}

\usepackage{graphicx}
\usepackage{caption}
\usepackage{subfig}
\usepackage{float}
\usepackage{color}
\usepackage{epstopdf}
\usepackage{booktabs}

\usepackage{appendix}
\usepackage{enumerate}
\usepackage{cleveref}
\usepackage{url}
\usepackage{mathtools}
\def\fnl{f_{\rm NL}}

\def\gnl{g_{\rm NL}}

\def\q{\mathbf{q}}
\def\v{\mathbf{v}}
\def\x{\mathbf{x}}

\def\k{\mathbf{k}}

\def\H{\mathcal{H}}
\def\ps{\mathbf{\Psi}}

\def\na{\mathbf{\nabla}}
\def\vp{\varphi}
\newcommand{\vs}{\varphi_{\G,s}}
\newcommand{\vl}{\varphi_{\G,l}}

\newcommand{\kd}{\kappa_2}

\def\kt{\kappa_3}

\def\dkt{\frac{d\kt/dM}{d\ln\sigma^{-1}/dM}}

\def\au{\alpha(k_1)}
\def\ad{\alpha(k_2)}
\def\at{\alpha(k_3)}
\def\ai{\alpha(k_i)}
\def\aj{\alpha(k_j)}

\newcommand{\MA}{{\rm M}}

\def\sigA{\sigma^2_\alpha}
\def\sigPNGU{\sigma^2_{\text{nG},1}}
\def\sigPNGD{\sigma^2_{\text{nG},2}}

\newcommand{\vk}{\vec{k}}

\newcommand{\dqc}{\frac{\derivd^3p}{(2\pi)^3}}
\newcommand{\dqcp}{\frac{\derivd^3p'}{(2\pi)^3}}
\newcommand{\dqcpp}{\frac{\derivd^3p''}{(2\pi)^3}}
\newcommand{\be}{\begin{equation}}
\newcommand{\ee}{\end{equation}}
\newcommand{\barr}{\begin{align}}
\newcommand{\earr}{\end{align}}

\newcommand{\cyc}{\ \text{cyc.}}

\newcommand{\Mpc}{\ \text{Mpc}}

\newcommand{\derivd}{\,\mathrm{d}} 

\newcommand{\degs}{\ \text{deg}^2}

\newcommand{\lin}{{\rm lin}}

\newcommand{\ini}{{\rm in}}

\newcommand{\G}{{\rm G}}

\newcommand{\E}{{\rm E}}
\newcommand{\La}{{\rm L}}
\newcommand{\LV}{{\rm LV}}
\newcommand{\PS}{{\rm PS}}
\newcommand{\ST}{{\rm ST}}

\newcommand{\R}{{\rm R}}
\newcommand{\p}{\mathbf{p}}
\newcommand{\INT}{{\int \frac{d\k_1}{(2\pi)^3} \int \frac{d\k_2}{(2\pi)^3} \, \delta^D (\k-\k_1-\k_2)}}
\newcommand{\INTn}{{\sum_{n=1}^{\infty}\int \frac{d\k_1}{(2\pi)^3} \ldots \int \frac{d\k_{n-1}}{(2\pi)^3} \int d\k_n \, \delta^D (\k- \ldots -\k_n)}}

\title{Galaxy bispectrum, primordial non-Gaussianity and redshift space distortions}

\author[a]{Matteo Tellarini,}
\author[a,b]{Ashley J. Ross,}
\author[c]{Gianmassimo Tasinato}
\author[a]{and David Wands}

\emailAdd{matteo.tellarini@port.ac.uk}
\emailAdd{ross.1333@osu.edu}
\emailAdd{g.tasinato@swansea.ac.uk}
\emailAdd{david.wands@port.ac.uk}

\affiliation[a]{Institute of Cosmology \& Gravitation, University of Portsmouth, Dennis Sciama Building, Portsmouth, PO1 3FX, United Kingdom}
\affiliation[b]{Center for Cosmology \& AstroParticle Physics, The Ohio State University, Columbus, OH 43210, USA}
\affiliation[c]{Department of Physics, Swansea University, Swansea, SA2 8PP, UK}

\abstract{Measurements of the non-Gaussianity of the primordial density field have the power to considerably improve our understanding of the physics of inflation. Indeed, if we can increase the precision of current measurements by an order of magnitude, a null-detection would rule out many classes of scenarios for generating primordial fluctuations. Large-scale galaxy redshift surveys represent experiments that hold the promise to realise this goal. Thus, we model the galaxy bispectrum and forecast the accuracy with which it will probe the parameter $\fnl$, which represents the degree of primordial local-type non Gaussianity. Specifically, we address the problem of modelling redshift space distortions (RSD) in the tree-level galaxy bispectrum including $\fnl$. We find novel contributions associated with RSD, with the characteristic large scale amplification induced by local-type non-Gaussianity. These RSD effects must be properly accounted for in order to obtain un-biased measurements of $\fnl$ from the galaxy bispectrum. We propose an analytic template for the monopole which can be used to fit against data on large scales, extending models used in the recent measurements.
Finally, we perform idealised forecasts on $\sigma_{\fnl}$ -- the accuracy of the determination of local non-linear parameter $\fnl$ -- from measurements of the galaxy bispectrum. 
Our findings suggest that current surveys can in principle provide $\fnl$ constraints competitive with \emph{Planck}, and future surveys could improve them further.
}

\keywords{Primordial non-Gaussianity, Large-Scale Structure}

\arxivnumber{1603.06814}

\begin{document}
\maketitle
\section{Introduction}\label{sec:intro}
Inflation in the very early universe is a simple mechanism for generating primordial density fluctuations from vacuum fluctuations. This gives rise to anisotropies in the cosmic microwave background (CMB) and the large-scale structures that we observe in the Universe today \cite{Liddle:2000cg}.
Different realizations of inflation lead to distinctive observational consequences, which can be studied through the predicted statistics for the primordial fluctuations. An important example is the non-Gaussian features associated with the non-vanishing $n$-point correlations functions $\langle \Phi_\ini (\k_1) \ldots \Phi_\ini (\k_n)$ of the primordial gravitational potential $\Phi_\ini(\x)$. Reconstructing the initial non-Gaussian pattern of primordial fluctuations from inflation out of late-time observations is a major focus of current cosmology research.

Inflationary mechanisms can lead to primordial non-Gaussian features in different ways \cite{Bartolo:2004if}: the most studied example is {\it local} non-Gaussianity, in which fluctuatuations 
  of the primordial potential $\Phi_\ini(\x)$  can be expressed as a power series of a single, Gaussian field $\vp_\G(\x)$ so that \cite{Gangui:1993tt,Verde:1999ij,Komatsu:2001rj}
\begin{equation}\label{eq:png}
 \Phi_\ini (\x) = \vp_\G (\x) + \fnl \left( \vp^2_\G (\x) - \langle \vp^2_\G \rangle \right) + \ldots\,,
\end{equation}
with a constant non-linearity parameter $\fnl$ controlling the deviations from purely Gaussian statistics.
In this work, we will focus our attention on local non-Gaussianity.

Local non-Gaussianity in real space leads to a primordial bispectrum for $\Phi_\ini$ (the Fourier transform of the 3-point function) which peaks 
 in the squeezed configuration ($k_1 \simeq k_2 \gg k_3$) in Fourier space. Scenarios leading to local non-Gaussianity -- and $\fnl$ greater than or of order
 one -- include multiple field inflationary models, or set-ups involving conversion mechanisms as curvaton or modulated reheating (see e.g.
  \cite{Bassett:2005xm} for a review). Single field, slow-roll
 inflation instead predicts an extremely small non-Gaussian signal in the squeezed configuration of the bispectrum, whose size is of the order of the tilt of the power spectrum \cite{Maldacena:2002vr}.
  
The state-of-the-art for measurements of $\fnl$ is given by \emph{Planck} satellite measurements of the CMB bispectrum \cite{Ade:2015ava} which gives $-9.2<\fnl<10.8$ at $95\%$ CL. Distinguishing $|\fnl|  \ll 1$ from $|\fnl| \sim 1$ is a key target to observationally  distinguish  single-field, slow-roll inflation from other scenarios \cite{Alvarez:2014vva}. However, CMB data alone may not be able to reach this goal, limited on large-scales by cosmic variance and on small scales by Silk damping. 

A promising possibility comes from the statistics of large-scale structure (LSS), and was first pointed out in the pioneering paper by Dalal {\em et al} \cite{Dalal:2007cu}. The coupling between the long and short modes present in local-type non-Gaussianity introduces a scale-dependent relation between dark matter halos and the underlying matter distribution, which scales  as $\fnl / k^2$ and therefore affects the  LSS power spectrum  \cite{Dalal:2007cu,Matarrese:2008nc,Slosar:2008hx,Afshordi:2008ru}. This contribution, induced by primordial non-Gaussianity (PNG), is amplified at large scales;
the clustering of LSS objects thus offers a way to measure the non-linear parameter $\fnl$ which is complementary to the CMB. By analyzing the galaxy power spectrum, constraints have already been set \cite{Slosar:2008hx,Xia1:2010,Xia2:2010,Ross:2013,Karagiannis:2013xea,Giannantonio:2013uqa,Ho:2013lda,Giannantonio:2013kqa,Leistedt:2014zqa} which are competitive with \emph{WMAP} \cite{WMAP9}, while future redshift surveys are expected to give results more stringent than \emph{Planck} \cite{Giannantonio:2011,Font-Ribera:2013rwa,Raccanelli:2014kga,Camera:2014bwa}. Moreover, the use of multi-tracer techniques is expected to allow constraints better than the naive cosmic variance limit \cite{Seljak:2008xr,Abramo:2013awa}. Such techniques are promising for further improving the bounds on $\fnl$ \cite{Yamauchi:2014ioa,Ferramacho:2014pua,Alonso:2015sfa,Fonseca:2015laa}.

Studies of the galaxy bispectrum indicate that an accuracy in determining local $\fnl$ of order
$\sigma_{\fnl} \sim \, \text{few}$ is achievable \cite{Scoccimarro:2003wn,Sefusatti:2007ih,Baldauf:2010vn}. The accuracy achievable could even be less than one if the survey is optimised for detecting PNG \cite{Dore:2014cca}. Considering such higher-order statistics gives access to the full shape information of the non-Gaussian signal, with the primordial one having a scale dependence even stronger than $k^{-2}$ \cite{Jeong:2009,Baldauf:2010vn,Tasinato:2013vna}. Additionally, information contained in the bispectrum potentially allows us to break the degeneracy in the power spectrum between $\fnl$ and the next-order non-Gaussian parameter $\gnl$ \cite{Roth:2012}, as discussed in \cite{Jeong:2009,Tasinato:2013vna}.

On the other hand, there are significant challenges in measuring $\fnl$ with LSS that are both theoretical and observational. 
At the theoretical level, as the Universe evolves density perturbations undergo non-linear evolution through gravitational collapse, and therefore we require an accurate modelling of the density evolution, capable of separating the primordial non-Gaussian signal from the one generated by clustering. Moreover, a precise description of how dark matter halos form starting from the primordial  density field is necessary.
Further, the high accuracy required for measuring a signal of $\fnl \sim 1$ implies that General Relativity (GR) effects cannot be neglected. 
Although in the simplest halo models they do not give rise to a scale-dependent bias \cite{Dai:2015jaa,dePutter:2015vga,Bartolo:2015qva}, GR effects do source contributions to the squeezed limit of the matter bispectrum \cite{Bartolo:2005xa,Tram:2016cpy} and generate secondary non-Gaussianities along the path of the photons from the emitting galaxy and the observer, in analogy with the CMB (see for instance \cite{Camera:2014bwa,DiDio:2015bua} and references therein).

At the observational level, several issues should be taken
into account, such as mask geometry and systematic effects, which can mimic the  scale dependence of PNG \cite{Huterer2013,Pullen2013,Agarwal:2013ajb}.
Moreover, redshift space distortions (RSD) are an additional source of complexity \cite{Hamilton:1997zq}: since the redshift measurements used to infer the distances of galaxies are contaminated by peculiar velocities, distortions appear along the line of sight. They can either be due to the in-fall of galaxies into clusters or due to the velocity dispersion inside a cluster, when its non-linear structure is resolved. The former leads to an apparent squashing of the clustering along the line of sight on large scales (say $k \lesssim 0.1 \, h/\Mpc$), modelled at linear level through the \emph{Kaiser factor} \cite{Kaiser:1987qv}, while the latter is responsible for elongation on small scales (say $k \gtrsim 0.1 \, h/\Mpc$), usually referred to as \emph{Fingers of God} (FoG)\footnote{Historically, the presence of FoG was recognised for the first time in \cite{1972MNRAS.156P...1J}.}.

In this work we address the problem of computing the galaxy bispectrum in redshift space with PNG of local type. For the purpose of obtaining an analytic result, we focus mainly on  large-scale regimes. We point out new potentially significant effects, induced by primordial non-Gaussianity, associated with large-scale amplifications of RSD. By decomposing the line of sight dependence of the bispectrum into spherical harmonics, we also make a prediction for the galaxy monopole, motivated by the recent measurement of \cite{Gil-Marin:2014sta}.

We also use a Fisher matrix analysis to estimate $\sigma_{\fnl}$ expected from the bispectrum of BOSS \cite{BOSS}, eBOSS \cite{Dawson:2015wdb}, DESI \cite{DESI}, Euclid \cite{EUCLID}. These results are approximate, but can be compared to forecast constraint from the power spectrum \cite{Font-Ribera:2013rwa,Zhao:2015gua} in order to estimate the potential power of bispectrum measurements in future. A comparison suggests that the bispectrum may improve the measurement of $\fnl$ by about an order of magnitude respect to the power spectrum of a single tracer. 

\section{Basics}\label{sec:basics}

In this section we briefly discuss  how the local non-linear parameter $\fnl$ affects the evolution of the density field and the formation of structures. We review our findings of  \cite{Tellarini:2015faa} and we define our notation. See also \cite{Bernardeau:2001qr} for a detailed review on these topics.

We concentrate on local type non-Gaussianity of the form given in \cref{eq:png}.  Local non-gaussianity induces correlations among fluctuations of different wave numbers, in particular between long and short modes. This becomes  apparent when taking the Fourier transform of  \cref{eq:png}: the second term proportional to $\fnl$ becomes a convolution, which couples different wavenumbers of the Gaussian mode  $\varphi_G$. This fact is important for our discussion. 
   
\subsection{Perturbation theory}\label{sec:pt}
The evolution equations for perturbations of a   cosmic fluid in an expanding Friedman-Robertson-Walker universe can be  formulated in terms of the matter overdensity $\delta(\x)$ and the corresponding velocity divergence  $\theta(x) = \nabla\cdot\v(\x)$, and then solved in perturbation theory by \cite{Bernardeau:2001qr}
\begin{align}
 \delta(\k,z) = &\INTn \mathcal{F}_n(\k_1,\ldots,\k_n,z)\delta_\lin(\k_1,z)\ldots\delta_{\lin}(\k_n,z) \label{eq:delta}\\
  \theta(\k,z) = &-f \H \INTn \mathcal{G}_n(\k_1,\dots,\k_n,z) \times \nonumber \\ & \times \delta_\lin(\k_1,z)\dots\delta_{\lin}(\k_n,z) \label{eq:theta}\,,
\end{align}
where $\H$ is the conformal Hubble parameter at redshift $z$, related to the Hubble parameter $H$ by $\H(z)=a H$, and $f$ is the logarithmic derivative of the linear growth factor $D(z)$, $f= d\ln D/d\ln a \approx \Omega_m^{4/7}(z)$ in $\Lambda$CDM \cite{1991MNRAS.251..128L}. In the matter-era $D(z) \propto (1+z)^{-1}$  and we use the normalization $D(0)=1$. The linearly evolving density field $\delta_{\lin}$ is related to the primordial gravitational potential through
\begin{equation}
 \label{eq:alpha}
 \delta_\lin (\k,z) = \alpha(k,z) \Phi_\ini (\k) \,,
\end{equation}
where the function $\alpha(k,z)$ is defined as
\begin{equation}
\label{eq:Phi}
\alpha(k,z) \equiv  \frac{2 k^2 c^2 T(k) D(z)}{3 \Omega_m H_0^2} \,.
\end{equation}
$T(k)$ is the transfer function, which goes to one as $k \rightarrow 0$. 
Note that the linearly evolving density (\cref{eq:alpha}) includes non-Gaussian terms in the presence of PNG. Thus
it is  useful to define the Gaussian part of the linearly evolving density field as 
\begin{equation}
 \label{eq:deltaG}
 \delta_{\G} (\k,z) = \alpha (k,z) \vp_\G (\k) \,.
\end{equation}

In general, the kernels $\mathcal{F}_n$ and $\mathcal{G}_n$ are time dependent. However, since they are  weakly sensitive to the underlying cosmology, we can compute
them  for an Einstein-De Sitter universe, where they are constant in time. At linear order they read $\mathcal{F}_1(\k) = \mathcal{G}_1(\k) = 1$, so that $\delta^{(1)} = \delta_{\G}$ and $\theta^{(1)} = - f \H \,\delta_{\G}$. 
The second-order solutions are  \cite{Bernardeau:2001qr}
\begin{align}
 \delta^{(2)}(\k,z) = & \INT \left[ \mathcal{F}_2(\k_1,\k_2) + \fnl \frac{\alpha(k)}{\alpha(k_1)\alpha(k_2)}\right] \delta_\G(\k_1,z) \delta_\G(\k_2,z) \label{eq:delta2}\\
 \theta^{(2)}(\k,z) = &- f \H \INT \times \nonumber \\ \qquad \qquad \qquad & {} \times \left[ \mathcal{G}_2(\k_1,\k_2) + \fnl \frac{\alpha(k)}{\alpha(k_1)\alpha(k_2)}\right]\delta_\G(\k_1,z)\delta_\G(\k_2,z) \,\label{eq:feta2}
\end{align}
with the kernels defined as 
\begin{align}
 \mathcal{F}_2(\k_1,\k_2) &= \frac{5}{7} + \frac{1}{2} \frac{\k_1\cdot\k_2}{k_1 k_2}\left(\frac{k_1}{k_2}+\frac{k_2}{k_1} \right) + \frac{2}{7} \frac{(\k_1\cdot\k_2)^2}{k_1^2 k_2^2} \,,\\
 \mathcal{G}_2(\k_1,\k_2) &= \frac{3}{7} + \frac{1}{2} \frac{\k_1\cdot\k_2}{k_1 k_2}\left(\frac{k_1}{k_2}+\frac{k_2}{k_1} \right) + \frac{4}{7} \frac{(\k_1\cdot\k_2)^2}{k_1^2 k_2^2} \,.
\end{align}
Notice that the couplings between modes of different wavelengths introduce a dependence on $\fnl$  in the second order solutions (\cref{eq:delta2,eq:feta2}).

The quantities $\delta(\bf x)$  and $\theta(\bf x)$ are expressed in Eulerian frame, with the initial 
spatial coordinate ${\bf q}$ in the Lagrangian frame being related to the evolved Eulerian coordinate ${\bf x}$ through the
formula
\be
{\bf x}( \bf q, \tau)\,=\,{\bf q}+\Psi(\bf q, \tau)\,,
\ee
where $\bf \Psi$ is the displacement field. 
Such relation is useful for obtaining  
an alternative way to write the second-order solution of \cref{eq:delta2}, which  will be needed in \cref{sec:ebivariate}.
 It  is given by \cite{peebles1980large,1992ApJ...394L...5B}:
\begin{align} 
\label{eq:secE}
\delta^{(2)}(\x,\tau) & = \frac{17}{21} (\delta_\lin (\x,z))^2 + \frac{2}{7} s^2(\x,z) - \ps(\x,z) \cdot \na\delta(\x,z) \,,
\end{align}
where $s^2=s_{ij}s^{ij}$ and $s_{ij}$ is the trace-free \emph{tidal tensor}, defined as
\begin{equation}
 s_{ij} \equiv \left( \nabla_i\nabla_j -  \frac{1}{3} \delta^K_{ij} \nabla^2 \right) \nabla^{-2}\delta \,,
\end{equation} 
and $\delta^K_{ij}$ is the Kronecker delta.
From now on, in order to simplify our expressions, we will not explicitly write the redshift dependence in the density and velocity fields.

\subsection{Press-Schechter approach}\label{sec:ps}
The Press-Schechter approach \cite{1974ApJ...187..425P} and its extensions \cite{Bardeen:1986,Bond:1991} provide a
 consistent framework for describing  the full non-linearly evolved density field, and in particular the
number of gravitationally collapsed dark matter halos, in terms of the initial, linearly growing density field (see \cite{Zentner:2006vw} for a pedagogical review). 
Dark matter halos are identified as peaks in the linearly growing density field of \cref{eq:alpha}, exceeding a suitable threshold value: this is usually assumed to be the linearly growing density amplitude for a spherically collapsed object, $\delta_c \simeq 1.686$ \cite{Liddle:2000cg}. 

The number density of objects with mass $\MA$ at redshift $z$ is called the \emph{mass function}. Defining  $\nu \equiv \delta_c / \sigma $, the Press-Schechter approach predicts the mass function to have the form
\begin{equation}
 \label{eq:nh}
  n_g (\MA,z) = f(\nu) \frac{\rho_m}{\MA} \biggr\vert \frac{d \ln \sigma}{d \MA} \biggr\vert \,,
\end{equation}
where the variance of the smoothed linearly-evolved density field is
\begin{equation}\label{eq:kd}
      \sigma^2 = \langle \delta_\lin^2 \rangle = \int \dqc \int \dqcp W_\MA(p) \alpha(p,z) W_\MA(p') \alpha(p',z) \langle \Phi_\ini(\p) \Phi_\ini(\p') \rangle \,.
  \end{equation}
We choose the window function $W_\MA (k,\R)$ to be the real-space top-hat filter of length scale $\R(\MA)=(3\MA/4 \pi \rho_m)^{1/3}$, with Fourier transform
\begin{equation}
 W_\MA (k,\R) = \frac{3}{(k \R)^3} \left[ \sin (k \R) - k \R \cos(k \R) \right] \,.
\end{equation}

In this work we follow \cite{Baldauf:2010vn} and assume the mass function to be a combination of the Press-Schechter $f_\PS$ \cite{1974ApJ...187..425P}, Sheth-Tormen $f_\ST$ \cite{Sheth:1999,Sheth:2001,Sheth:2002} and Lo Verde $f_\LV$ \cite{LoVerde:2007ri} mass functions:
\begin{equation}\label{eq:stlv}
 f (\nu) = f_\ST \, \frac{f_\LV}{f_\PS} = f_\ST (\nu) \left[ 1 + \frac{1}{6}\left(\kt(\MA) H_3(\nu) - \frac{d\kt(\MA)/d\MA}{d\ln\sigma^{-1}/d\MA} \frac{H_2(\nu)}{\nu} \right) \right] \,,
\end{equation}
where $H_n$ is the $n$-th Hermite polynomial and the $3$rd cumulant $\kt(\MA)$ is approximately \cite{2011JCAP...08..003L}
\begin{align}
  \label{eq:kt}
  \kt (\MA) & \approx \, \fnl \left( 6.6 \times 10^{-4} \right) \left[ 1 - 0.016 \ln \left( \frac{\MA}{h^{-1} \MA_{\odot}} \right) \right] \,.
\end{align}
\Cref{eq:stlv} is assumed to be a reasonable description of objects formed from ellipsoidal collapse with non-Gaussian (nG) initial conditions. Further details on the mass function can be found in \cref{app:massfunction}.

\subsection{The bivariate model}\label{sec:bivariate}

The fact that primordial non-Gaussianity couples modes of different wavenumber has important consequences for the Press-Schechter theory of structure
formation. 
Since dark matter halos are associated with peaks of the matter density contrast, we 
 can expect that the distribution of halo overdensity  depends not only on the local value of the matter density contrast, but also on how it is distributed around a given position.
  This information is encoded in the 
 correlation functions of the matter density contrast, that in turn are affected by PNG. A convenient way for describing this fact  introduces
  the concept of  peak-background split (PBS), and  leads to important consequences when studying the overdensity of halos with PNG of local-type \cite{Dalal:2007cu,Slosar:2008hx,Giannantonio:2009ak,Baldauf:2010vn,Smith:2011ub,Tasinato:2013vna,Tellarini:2015faa}. The PBS approach identifies the primordial Gaussian field $\vp_\G$ as a superposition of statistically independent long and short modes
\begin{equation}
 \label{eq:pbs}
\vp_\G(\q) = \vl(\q) + \vs(\q) \,, 
\end{equation}
where the use of $\q$ indicates that we work with the initial (or linearly evolved) quantities, therefore in the Lagrangian frame.

The (arbitrary) length scale $l$ is chosen such that the short modes are responsible for the collapse of matter into objects on a scale $R \ll l$, while the long modes only perturb the approximately homogeneous background cosmology. 
 This in turn  requires the variable $\nu$ in the mass function  to be replaced with a {\it local} value
\begin{equation}
\label{eq:nul}
  \nu = \frac{\delta_c}{\sigma} \, \longrightarrow \, \nu(\q) = \frac{\delta_c - \delta_{\lin,l}(\q)}{\sigma_l (\q)} \,,
\end{equation}
where the long density modes are defined as
\begin{equation}
  \delta_{\lin,l} (\k) = \delta_{\G,l} + \fnl \alpha \left( \vl^2 - \langle \vl^2 \rangle \right)
\end{equation}
and the effective variance of the short modes is now modulated, through PNG,  by $\vl$,
\begin{equation}
 \label{eq:sigmal}
  \sigma_l = (1 + 2 \fnl \vl) \sigma \,.
\end{equation}
By treating $\delta_{\lin,l}$ and $\sigma_l$ as  independent perturbations controlling the halo overdensity
 in Lagrangian space, $ \delta_g^\La (\q)$, this quantity can be then conveniently  expressed  as a double Taylor expansion, 
\begin{equation}
 \label{eq:bivariate}
 \delta_g^\La (\q) = \frac{n_g(\q) - \langle n_g \rangle}{\langle n_h \rangle} = \, b^\La_{10} \delta_{\lin} + b^\La_{01} \vp_\G + b^\La_{20} (\delta_{\lin})^2 + b^\La_{11} \delta_{\lin} \vp_\G + b^\La_{02} \vp_\G^2 + \dots\, ,
\end{equation}
where we dropped the subscript $l$ in the right hand side to simplify the notation; the Lagrangian bias coefficients $b_{ij}^\La$ derived from the mass function of \cref{eq:stlv} are quoted in \cref{app:massfunction}. \Cref{eq:bivariate} is known as the \emph{bivariate model} for the halo/galaxy overdensity \cite{Giannantonio:2009ak} (see
also \cite{Assassi:2015fma} for a recent application of a bivariate model for an effective field theory approach to galaxy biasing).

\subsection{Galaxy overdensity in the Eulerian frame}\label{sec:ebivariate}
The bivariate model describes the statistics of the objects in the Lagrangian frame, while their dynamics are obtained by the transformation to Eulerian coordinates. 
As we explained,  the two frames are linked by $\x(\q,\tau)=\q+\ps(\q,\tau)$, where the displacement field $\ps$ is the dynamical quantity in the Lagrangian picture. The transformation can be performed under the conservation of the number density of objects in a given volume; if there is no velocity bias between objects and matter, then \cite{Catelan:1997qw}
\begin{equation}
\label{eq:nhE}
1+\delta_g^\E (\x,z) = [1+\delta(\x,z)] \left[ 1+ \delta_g^\La (\q,z) \right] \,.
\end{equation} 

In \cite{Giannantonio:2009ak,Baldauf:2010vn}, the halo/galaxy overdensity in the Eulerian frame is obtained by assuming spherical collapse. By dropping this assumption, 
we get 
a more general (non-local) result   \cite{Tellarini:2015faa}:
\begin{equation}
\label{eq:deltahk}
\begin{split}
\delta_g^\E(\k)=\,& b_{10}^\E \delta + b_{01}^\E \vp_\G + b_{20}^\E \delta \ast \delta + b_{11}^\E \delta \ast \vp_\G + b_{02}^\E \vp_\G \ast \vp_\G - \frac{2}{7} b_{10}^\La s^2 - b_{01} n^2 \,,
\end{split}
\end{equation}
where $\ast$ stands for a convolution and the Eulerian bias coefficients are related to Lagrangian ones through
\begin{equation}\label{eq:eul_bias}
\begin{split}
 b_{10}^\E & = 1+b_{10}^\La \,, \\
 b_{01}^\E & = b_{01}^\La  \,, \\
 b_{20}^\E & = \frac{8}{21} b_{10}^\La + b_{20}^\La \,, \\
 b_{11}^\E & = b_{01}^\La + b_{11}^\La \,, \\
 b_{02}^\E & = b_{02}^\La \,.
\end{split}
\end{equation}
From now on, we drop the superscript $\E$ to indicate the Eulerian bias coefficients, in order to simplify the notation. However, we  continue to write the superscript $\La$ where needed for avoiding confusion. As discussed in \cref{sec:pt}, the Fourier transform of the density field in Eulerian coordinates is
\begin{equation}
\label{eq:deltaEk}
\delta (\k)  = \, \delta_\G(\k) +   \int \frac{d\q}{(2 \pi)^3} \left[ \mathcal{F}_2(\q,\k-\q) + \fnl \frac{\alpha(k)}{\alpha(q) \alpha(\vert \k-\q \vert)} \right] \delta_\G(\q)\delta_\G(\k-\q) \,,
\end{equation}
while the tidal term $s^2$ reads 
\cite{Catelan:2000vn,Baldauf:2012hs}
\begin{equation}
 s^2 (\k) = \, \int \frac{d\q}{(2 \pi)^3} \mathcal{S}_2(\q,\k-\q)\delta_\G(\q)\delta_\G(\k-\q) \,,
\end{equation}
with the kernel defined as follows
\begin{equation}
 \mathcal{S}_2(\k_1,\k_2) = \, \frac{\left( \k_1 \cdot \k_2 \right)^2}{k_1^2 k_2^2} - \frac{1}{3} \,.
\end{equation}
The non-Gaussian shift term $n^2$ is a consequence of the displacement of halos/galaxies respect to their initial positions $\q$ in the Lagrangian frame, which affects the field $\vp_\G (\x)$:
\begin{equation}
 n^2 (\k) = \, 2 \int \frac{d\q}{(2 \pi)^3} \mathcal{N}_2(\q,\k-\q)\frac{\delta_\G(\q)\delta_\G(\k-\q)}{\alpha(\vert \k-\q \vert)}\,,
\end{equation}
where the kernel is
\begin{equation}
 \mathcal{N}_2(\k_1,\k_2) = \, \frac{\k_1 \cdot \k_2}{2 k_1^2} \,.
\end{equation}

\noindent
We use the following standard
  definitions for the galaxy power spectrum $P_{gg}$ and bispectrum $B_{ggg}$:
\begin{align*}
 \langle \delta^\E_{g}(\k_1) \delta^\E_{g}(\k_2) \rangle &= (2\pi)^3 \delta^D(\k_1 + \k_2 ) P_{gg}(\k_1) \,,\\
 \langle \delta^\E_{g}(\k_1) \delta^\E_{g}(\k_2) \delta^\E_{g}(\k_3) \rangle &= (2\pi)^3 \delta^D(\k_1 + \k_2 + \k_3) B_{ggg}(\k_1,\k_2,\k_3) \,.
\end{align*}

At   tree-level, they  can be conveniently written as
\begin{align}
 P_{gg}(\k_1) & = E_1^2(\k_1) P(k_1) \label{eq:pgg} \,,\\
 B_{ggg}(\k_1,\k_2,\k_3) & = 2 E_1(\k_1) E_1(\k_2) E_2(\k_1,\k_2) P(k_1) P(k_2) + 2\cyc \label{eq:bggg}\,,
\end{align}

\noindent
where $P(k)$ is the matter power spectrum for the Gaussian source field $\varphi_G$, while
  the kernels $E_i$ are defined as
\begin{align}
  E_1 (\k_1) = \, & b_{10} + \frac{b_{01}}{\alpha(k_1)} \\
  \label{expE2}
  E_2 (\k_1,\k_2) =\, & b_{10} \left[F_2(\k_1,\k_2) + \fnl \frac{\alpha(\vert \k_1 + \k_2 \vert)}{\alpha(k_1)\alpha(k_2)} \right] + \left[b_{20} - \frac{2}{7} b_{10}^\La S_2(\k_1,\k_2)\right] \\
	  & + \frac{b_{11}}{2} \left[\frac{1}{\alpha(k_1)}+\frac{1}{\alpha(k_2)}\right] + \frac{b_{02}}{\alpha(k_1)\alpha(k_2)} - b_{01} \left[ \frac{N_2(\k_1,\k_2)}{\alpha(k_2)}+\frac{N_2(\k_2,\k_1)}{\alpha(k_1)}\right] \,. \nonumber
\end{align}
The term $b_{01}/\alpha(k_1) \propto \fnl/k^2_1$ in $E_1$ is the so-called scale-dependent bias: it is responsible for deviations on the large-scale clustering, with respect to the scale-independent bias $b_{10}$. The first and second term that appears in $E_2$ account for non-linear clustering and non-linear biasing respectively, while the terms in the bottom line describe the non-linear effects due to PNG. A detailed analysis of the results for the bispectrum  is presented in \cite{Tellarini:2015faa}.

To conclude this section, let us point out that for
avoiding   the complexities of a full bispectrum measurement, a new observable is proposed in \cite{Chiang:2014oga} in terms
of position dependent power spectrum. We explore this possibility  in \cref{app:pdps}, in light of the result of \cref{eq:bggg}.
\section{Redshift space distortions}\label{sec:rsd}
The peculiar velocities of galaxies contaminate the redshift measurements of surveys, resulting in distortions along the line of sight. The in-fall of galaxies into clusters is responsible for large-scale distortions, while the velocity dispersion inside a cluster leads to the Fingers of God (FoG), usually a small-scale effect. In this paper, for the first time, we study how PNG affects the bivariate halo distribution when formulated in redshift space, finding new and potentially sizeable large scale effects. We model RSD within a perturbative approach and focus mainly on the large scale effects, for the purpose of obtaining analytic results.
  
An object at some position $\x$ and redshift $z$  appears in redshift space at position \cite{Matsubara:1999du}:
\begin{align}
\x_s(z) = & ~ \x + (1+z) \frac{\v(\x) \cdot \hat{x}}{H(z)} \hat{x} = \x +\frac{v_x(\x)}{\H(z)}\hat{x} 
\nonumber\\
\label{eq:rsdmapping}
\approx&~ \x + f u_z(\x) \hat{z} ~.
\end{align}
In the second line  we use $f\,\equiv\,d \ln{ D}/d \ln{a}$, and we introduce the reduced component along the line of sight, $u_z$, given by
\begin{equation}\label{eq:reducedv}       
 u_z(\k)=\mathbf{u}(\k) \cdot \hat{z} = -\frac{i \mu}{k} \frac{\theta(\k)}{f\H} = \frac{i \mu}{k} \eta(\k) \,,
\end{equation}
where $\mu \equiv \hat{k} \cdot \hat{z} = \k \cdot \hat{z} / k$. 
The second line of \cref{eq:rsdmapping}  holds under the \emph{plane parallel} (or \emph{distant observer}, or \emph{flat sky}) approximation.
If the distances between galaxies are much smaller than the distance between the observer and the galaxies (resulting therefore in a small transverse component with  respect to the radial direction) the line of sight $\hat{x}$ can be assumed to be fixed along $\hat{z}$, pointing towards the centre of the galaxies of interest.
When used to compute the galaxy power spectrum, this approximation has been shown to be valid for pairs separated by an angle less than $10^\circ$ \cite{Szalay:1997cc}. If one has to use the full data from surveys such as BOSS and Euclid, where distances between galaxies can be comparable with the observer distance, then wide angle effects must be considered as well. This problem has been addressed in several works with various approaches (see for instance \cite{Heavens:1994iq,Tegmark:1994pa,1996ApJ...462...25Z,Hamilton:1995px,Szalay:1997cc,Szapudi:2004gh,Matsubara:2004fr,Papai:2008bd,Reimberg:2015jma}), investigated in numerical simulations (for example in \cite{2010MNRAS.409.1525R}) and the impact on measurements considered \cite{2012MNRAS.420.2102S,Yoo:2013zga}. 
However, in this first work about the effect of PNG on the bispectrum in redshift space, as predicted by the bivariate model, we will neglect for simplicity wide angle effects. Since we will show that PNG could be enhanced by RSD on large scales, in regimes where wide angle effects may not be negligible, the results of \cref{sec:rsd} should be considered as a zero order approximation of a more general framework which combines PNG, RSD and wide angle effects. The development of this framework is left for future work.

\Cref{eq:reducedv} approximates the redshift space mapping of \cref{eq:rsdmapping} as a power series. By comparing \cref{eq:reducedv} with \cref{eq:theta}, it follows that
\begin{equation}
 \eta(\k)= \sum_{n=1}^{\infty} \int \, \frac{d\k_1}{(2\pi)^3} \dots \int \, \frac{d\k_{n-1}}{(2\pi)^3} \int \, d\k_n \delta^D(\k-\k_1 - \ldots - \k_n) {\cal G}_n(\k_1,\dots,\k_n) \delta_{\lin}(\k_1) \dots \delta_{\lin}(\k_n) \,.
\end{equation}

\subsection{Galaxy overdensity in redshift space}\label{sec:deltags}
The transformation from real to redshift space is obtained by requiring the conservation of the number density of objects,
\begin{equation}
 [1+\delta_g^s(\x_s)]d\x_s = [1+\delta^\E_g(\x)]d\x\,,
\end{equation}
so that in Fourier space we have \cite{Scoccimarro:2004tg}
\begin{align}
  \delta_g^s(\k) & = \int \, d\x_s \delta_g^s(\x_s) e^{-i\k \cdot \x_s} \\
	       & = \int d\x \left[\delta^\E_g(\x)+1\right] e^{-i\k \cdot (\x+f u_z(\x)\hat{z})} - \int \, d\x e^{-i\k \cdot \x} \\
	       & = \delta^\E_g(\k) + \int \, d\x e^{-i\k \cdot \x}\left( e^{- if k_z u_z(\x)} - 1\right) \left[\delta^\E_g(\x)+1\right] \\
	       & = \delta^\E_g(\k) - \int \, d\x e^{-i\k \cdot \x}\left( i f k_z u_z(\x) +\frac{1}{2} f^2 k_z^2 u_z^2(\x) + \dots \right) \left[\delta^\E_g(\x)+1\right] \label{eq:smallu}\\
	       & = \delta^\E_g(\k) +f \mu^2 \eta(\k) - \int d\x e^{-i\k \cdot \x} \left(i f k_z u_z(\x) \delta_g(\x) + \frac{1}{2} f^2 k_z^2 u_z^2(\x) + \dots \right) \label{eq:fogng} \,,
 \end{align}
where \cref{eq:smallu} holds under the assumption of a small velocity component along the line of sight, $|u_z|\ll
1 $.

In general, the redshift-space galaxy overdensity can be written as
\begin{equation}\label{eq:dks}
 \delta_g^s(\k) = \sum_{n=1}^{\infty}  \int \, \frac{d\k_1}{(2\pi)^3} \dots \int \, \frac{d\k_{n-1}}{(2\pi)^3} \int \, d\k_n \delta^D(\k-\k_1 - \ldots - \k_n) Z_n(\k_1,\dots,\k_n) \delta_\G (\k_1) \dots \delta_\G (\k_n)
\end{equation}
where the redshift space kernels $Z_n(\k_1,\dots,\k_n)$ are,  up to second order 
\begin{align}
   Z_1 =\,& b_{10} \underbracket{\vphantom{\frac{b_{01}}{\alpha}} \left(1 + \beta \mu^2 \right)}_{\text{Kaiser}} + \underbracket{\frac{b_{01}}{\alpha}}_{\text{nG}1} \label{eq:z1kernel}\\
   Z_2 =\,& \underbracket{b_{10} \left[F_2(\k_1,\k_2) + \fnl \frac{\alpha(k)}{\alpha(k_1)\alpha(k_2)} \right]}_{\text{SQ}1} + \underbracket{\left[b_{20} - \frac{2}{7} b_{10}^\La S_2(\k_1,\k_2)\right]}_{\text{NLB}} \nonumber\\
	  &+ \underbracket{\frac{b_{11}}{2} \left[\frac{1}{\alpha(k_1)}+\frac{1}{\alpha(k_2)}\right] + \frac{b_{02}}{\alpha(k_1)\alpha(k_2)} - b_{01} \left[ \frac{N_2(\k_1,\k_2)}{\alpha(k_2)}+\frac{N_2(\k_2,\k_1)}{\alpha(k_1)}\right]}_{\text{nG}2} + \nonumber\\
	  &+\underbracket{ f \mu^2 \left[G_2(\k_1,\k_2) + \fnl \frac{\alpha(k)}{\alpha(k_1)\alpha(k_2)} \right]}_{\text{SQ}2} + \underbracket{\frac{f^2 k^2 \mu^2}{2} \frac{\mu_{1}\mu_{2}}{k_1 k_2}+b_{10} \frac{f \mu k}{2} \left(\frac{\mu_{1}}{k_1}+\frac{\mu_{2}}{k_2}\right)}_{\text{FOG}} +\nonumber\\
	  & + \underbracket{b_{01} \frac{f \mu k}{2} \left[ \frac{\mu_{1}}{k_1 \alpha(k_2)} + \frac{\mu_{2}}{k_2 \alpha(k_1)}\right]}_{\text{FoGnG}} \label{eq:z2kernel} ~,
\end{align}
with $\mu_i \equiv \hat{k}_i \cdot \hat{z} = \k_i \cdot \hat{z} / k_i$.
\noindent
These kernels $Z_i$ are important since they allow us to express the tree-level galaxy power spectrum and bispectrum in redshift space,
 by replacing $E_{1\,(2)} \rightarrow Z_{1\,(2)}$ in \cref{eq:pgg,eq:bggg}:
\begin{align}
 P^s_{gg}(\k_1) & = Z_1^2(\k_1) P(k_1) \label{eq:pggs} \\
 B^s_{ggg}(\k_1,\k_2,\k_3) & = 2 Z_1(\k_1) Z_1(\k_2) Z_2(\k_1,\k_2) P(k_1) P(k_2) + 2\cyc \label{eq:bgggs} \,,
\end{align}
including the effects of PNG. The resulting expression 
for $ B^s_{ggg}$ -- whose physics we will discuss more in detail in the next section -- represents one of the main results of this paper\footnote{A common factor $\mathcal{D}^B_{\text{FoG}}(k_1,k_2,k_3,\sigma^B_{\text{FoG}}[z])$ is usually included in the r.h.s of \cref{eq:bgggs}, accounting for the FoG damping due to intra-cluster velocity dispersion, beyond linear level \cite{Scoccimarro:1999ed}. This phenomenological extension, which describes N-body data, will not be considered here for the purpose of getting an analytic result in the next section.}.
At this stage we can make some considerations with respect to the various contributions to the $Z_i$, which we will then develop in what comes next.
  
Considering $Z_1$, it contains the bias $b_{10}$ and the scale-dependent correction (nG$1$). At linear level, RSD introduce the quantity $\beta \mu^2 \ge 0$, explaining why objects are more clustered in redshift space (compared to real space). The term $(1+\beta \mu^2)$ is often referred to as the `Kaiser factor' \cite{Kaiser:1987qv}, where $\beta = f/b_{10}$ and is regularly accounted for in studies of galaxy clustering (e.g.~\cite{Anderson14}). 

For $Z_2$, we notice that the contributions that we label with SQ$1$ (linear squashing), NLB (non-linear bias), nG$2$ (second order non Gaussian effects) are already present in the expressions for $E_2$ (\cref{expE2}) controlling the bispectrum in real space. On the other hand, the remaining three contributions are generated by redshift-space distortions. The quantities SQ$2$ (second order squashing) and FOG are already well studied in the literature. 
 
Interestingly, we notice the presence of a qualitatively new term (FoGnG), induced by PNG. It mimics the FOG contribution, {\it but} with an amplification of $1/\alpha(k)$. In a sense, it is an analogue for RSD of the scale-dependent bias of the galaxy power spectrum induced by PNG. The FoGnG term is sourced by the coupling between $u_z$ and $\vp_\G$ (first integrand in \cref{eq:fogng}), potentially affecting large-scale measurements. Therefore, neglecting it would introduce a systematic error, resulting in a biased $\fnl$ measurement from the tree-level of the bispectrum\footnote{At the power spectrum level, the FoGnG term enters as a loop correction. However, we do not consider its consequences in this paper, since we only focus on large scale effects.}.

As pointed out in \cite{Bernardeau:2001qr}, the galaxy overdensity of \cref{eq:dks} is the result of two approximations: one is the power series expansion (\cref{eq:reducedv}) of the redshift space mapping (\cref{eq:rsdmapping}), the other one is the perturbative expansion of $\delta(\k)$ and $\theta(\k)$ (\cref{eq:delta,eq:theta}). Therefore, the perturbation theory in redshift space is expected to break down on larger scales than in real space. However, replacing the kernels with effective kernels calibrated against simulations can extend the validity of the results based on \cref{eq:dks}, as shown in \cite{Gil-Marin:2014pva}. Although we will not implement these techniques  here, the replacement is straightforward.

\subsection{Galaxy bispectrum monopole}\label{sec:gmono}

In this section, we choose to investigate the galaxy bispectrum monopole, i.e.~the angle averaged bispectrum along the direction of the line of sight.
The reason is to extend the monopole model used in the recent measurement by Gil-Marin et al.~\cite{Gil-Marin:2014pva,Gil-Marin:2014sta,Gil-Marin:2014baa} to the case of PNG. The result of this section can thus be applied in a similar analysis, aiming to measure $\fnl$.

The galaxy bispectrum in redshift space is a function of five variables: three of them (say $k_1$, $k_2$ and $\hat{k_1} \cdot \hat{k_2} = \cos\theta_{12}$) fully define the shape of the triangle, while the polar angle $\omega=\arccos\mu_1$ and the azimuthal angle $\phi$ about $\hat{k}_1$ describe how it is oriented with respect to the line of sight. All the angles between the vectors $\k_1$, $\k_2$, $\k_3$ and the line of sight $\hat{z}$ can be written in terms of $\mu_1$ and $\phi$ \cite{Scoccimarro:1999ed}:
\begin{equation}
\mu_1=\cos\omega=\hat{k}_1\cdot\hat{z}\,,\qquad \mu_2=\mu_1 \cos\theta_{12}-\sqrt{(1-\mu_1^2)} \sin\theta_{12} \cos\phi\,,\qquad \mu_3=-\frac{k_1}{k_3}\mu_1-\frac{k_2}{k_3}\mu_2  \,.
\end{equation}
The $(\mu_1,\phi)$-dependence introduced by redshift space distortions can be conveniently decomposed into spherical harmonics,
\begin{equation}
 B_{ggg}^s(\k_1,\k_2,\omega,\phi) = \sum_{l=0}^{\infty} \sum_{m=-l}^{l} B_{ggg}^{s\,(l,m)}(\k_1,\k_2) Y_{lm}(\omega,\phi) \,.
\end{equation}
As we mentioned, we focus only on the monopole  $(l=0,m=0)$, i.e.~the average over all the possible orientations of the bispectrum with respect to the line of sight,
\begin{equation}
 B_{ggg}^{s\,(0,0)}(\k_1,\k_2)=\frac{1}{4\pi}\int_{-1}^{+1} d\mu_1 \int_{0}^{2\pi} d\phi B_{ggg}^s(\k_1,\k_2,\omega,\phi)\,,
\end{equation}
although the large-scale enhancement of PNG in redshift space, associated with the term called FoGnG (see \cref{eq:z2kernel}), is found also in higher multipoles, since it is not cancelled by angular integrations.

We start by quoting the monopole for Gaussian initial conditions ($\fnl = 0$) \cite{Scoccimarro:1999ed,Gil-Marin:2014pva}
\begin{equation}
\begin{split}
 B_{ggg}^{s\,G\,(0,0)}(\k_1,\k_2) =\,& b_{10}^4\Biggr\{ \frac{1}{b_{10}}\mathcal{F}_2 (\k_1,\k_2) \mathcal{D}_{\text{SQ1}} +\frac{1}{b_{10}}\mathcal{G}_2 (\k_1,\k_2)\mathcal{D}_{\text{SQ2}} + \\ & +\left[\frac{b_{20}}{b_{10}^2} - \frac{2}{7} \frac{b_{10}^\La}{b_{10}^2}\mathcal{S}_2(\k_1,\k_2)\right]\mathcal{D}_{\text{NLB}} + \mathcal{D}_{\text{FOG}}\Biggr\}P(k_1)P(k_2)+2\cyc \,.
\end{split}
\end{equation}
The terms $\mathcal{D}_{\text{SQ1}}$ and $\mathcal{D}_{\text{SQ2}}$ represent the linear and non-linear contributions to the large-scale squashing, $\mathcal{D}_{\text{NLB}}$ the non-linear bias contribution and, finally, $\mathcal{D}_{\text{FOG}}$ accounts for the linear part of FoG, i.e.~the damping effect due to velocity dispersion.
The labelling that we introduced in \cref{eq:z1kernel,eq:z2kernel} helps to understand where these factors come from: schematically $\mathcal{D}_{\text{SQ1}}$ is the result of the angular average of the Kaiser factor squared times the term SQ1, $\mathcal{D}_{\text{SQ2}}$ of the Kaiser factor squared times SQ2 and so on. \Cref{app:dfactors} further clarifies these points, with explicit expressions for the $\mathcal{D}$-factors. Here and after we omit their explicit dependence $\mathcal{D}(k_i,k_j,\cos\theta_{ij},y_{ij},\beta)$ to simplify the notation, but they are among the quantities to be permutated.

We now generalise the previous result to the case of local-type PNG; it can be written as
\begin{align}
 \label{eq:Bsggg0}
 B_{ggg}^{s\,(0,0)}(\k_1,\k_2) =\,& b_{10}^4\Biggr\{ \frac{1}{b_{10}}\biggr[ \mathcal{F}_2 (\k_1,\k_2) +\fnl\frac{\at}{\au\ad} \biggr] \mathcal{D}_{\text{SQ1}} \mathcal{R}_{\text{SQ1}} + \nonumber \\
  + & \frac{1}{b_{10}}\biggr[ \mathcal{G}_2 (\k_1,\k_2) +\fnl\frac{\at}{\au\ad} \biggr]\mathcal{D}_{\text{SQ2}}\mathcal{R}_{\text{SQ2}} + \nonumber \\ 
  & +\left[\frac{b_{20}}{b_{10}^2} - \frac{2}{7} \frac{b_{10}^\La}{b_{10}^2}\mathcal{S}_2(\k_1,\k_2)\right]\mathcal{D}_{\text{NLB}}\mathcal{R}_{\text{NLB}} + \mathcal{D}_{\text{FOG}} \mathcal{R}_{\text{FOG}} \\
  & + \frac{1}{b_{10}^2} \biggr[ \frac{b_{11}}{2}\left(\frac{1}{\au}+\frac{1}{\ad}\right) - b_{01} \left( \frac{\mathcal{N}_2(\k_1,\k_2)}{\ad} + \frac{\mathcal{N}_2(\k_2,\k_1)}{\au} \right) + \frac{b_{02}}{\au \ad}\biggr]\times \nonumber \\ & \times \mathcal{D}_{\text{nG}2}\mathcal{R}_{\text{nG}2} + \frac{b_{01}}{b_{10}}\mathcal{D}_{\text{FoGnG}} \mathcal{R}_{\text{FoGnG}} \Biggr\}P(k_1)P(k_2)+2\cyc \nonumber \,,
\end{align}
where we have introduced the correction factors
\begin{align}
 \mathcal{R}_{\text{SQ1}} = \,& 1 + \frac{1}{\mathcal{D}_{\text{SQ1}}}\left(\frac{b_{01}^2}{b_{10}^2} \frac{2}{\au \ad} + \frac{b_{01}}{b_{10}} \mathcal{D}_{\text{SQ}1}^{\text{nG}1} \right) \\
 \mathcal{R}_{\text{SQ2}} = \,& 1 + \frac{1}{\mathcal{D}_{\text{SQ2}}}\left( \frac{b_{01}}{b_{10}} \mathcal{D}_{\text{SQ}2}^{\text{nG}1} + \frac{b_{01}^2}{b_{10}^2} \mathcal{D}_{\text{SQ}2}^{\text{nG}1^2} \right) \\
 \mathcal{R}_{\text{NLB}} = \,& 1 + \frac{1}{\mathcal{D}_{\text{NLB}}}\left( \frac{b_{01}^2}{b_{10}^2} \frac{2}{\au \ad} + \frac{b_{01}}{b_{10}} \mathcal{D}_{\text{NLB}}^{\text{nG}1} \right) \\
 \mathcal{R}_{\text{FOG}} = \,& 1 + \frac{1}{\mathcal{D}_{\text{FOG}}}\left( \frac{b_{01}}{b_{10}} \mathcal{D}_{\text{FOG}}^{\text{nG}1} + \frac{b_{01}^2}{b_{10}^2} \mathcal{D}_{\text{FOG}}^{\text{nG}1^2}\right) \\
 \mathcal{R}_{\text{nG}2} = \,& 1 + \frac{1}{\mathcal{D}_{\text{nG}2}}\left( \frac{b_{01}^2}{b_{10}^2} \frac{2}{\au \ad} + \frac{b_{01}}{b_{10}} \mathcal{D}_{\text{nG}2}^{\text{nG}1} \right) \\
 \mathcal{R}_{\text{FoGnG}} = \,& 1 + \frac{1}{\mathcal{D}_{\text{FoGnG}}}\left( \frac{b_{01}}{b_{10}} \mathcal{D}_{\text{FoGnG}}^{\text{nG}1} + \frac{b_{01}^2}{b_{10}^2} \mathcal{D}_{\text{FoGnG}}^{\text{nG}1^2} \right) \,.
\end{align}
We see that PNG enters into the expression (\ref{eq:Bsggg0}) in four different ways:
\begin{itemize}
 \item The kernels $\mathcal{F}_2$ and $\mathcal{G}_2$ acquire a correction proportional to $\fnl$, as seen in \cref{sec:pt}.
 \item The linear (SQ1) and non-linear (SQ2) squashing, non-linear biasing (NLB) and linear part of FoG (FOG) are modified by the correction factors $\mathcal{R}_{\text{SQ1}}$, $\mathcal{R}_{\text{SQ2}}$, $\mathcal{R}_{\text{NLB}}$ and $\mathcal{R}_{\text{FOG}}$, respectively. In these, for instance,  $\mathcal{D}_{\text{SQ}1}^{\text{nG}1}$ comes from the angular average of the Kaiser factor times SQ1 times nG$1$ (linear effect of PNG), $\mathcal{D}_{\text{NLB}}^{\text{nG}1}$ from the Kaiser factor times NLB time nG$1$, $\mathcal{D}_{\text{SQ}2}^{\text{nG}1^2}$ from the SQ2 term times nG$1$ squared, and so on.
 \item  Non-Gaussianity distortions appear in the non-linear effect of PNG through the term $\mathcal{D}_{\text{nG}2}\mathcal{R}_{\text{nG}2}$. In particular, $\mathcal{D}_{\text{nG}2}$ is generated by the integration of the Kaiser factor squared times the non-linear effect of PNG (nG$2$) and $\mathcal{D}_{\text{nG}2}^{\text{nG}1}$ by the angular average of the Kaiser factor times nG$1$ times nG$2$.
 \item Importantly, a new set of terms appear, potentially relevant at large scales, related to the quantity called FoGnG in \cref{eq:z2kernel}. 
   $\mathcal{D}_{\text{FoGnG}}$ is the result of the integration of the Kaiser factor squared times FoGnG, $\mathcal{D}_{\text{FoGnG}}^{\text{nG}1}$ of the Kaiser factor times FoGnG times nG$1$ and, finally, $\mathcal{D}_{\text{FoGnG}}^{\text{nG}1^2}$ by the angular average of the FoGnG term times nG$1$ squared.
\end{itemize}
 Appendix \ref{app:dfactors} further discusses in detail  all the $\mathcal{D}$-factors that we schematically described here. 

\begin{figure}[htb]
\centering
\includegraphics[width=1\textwidth]{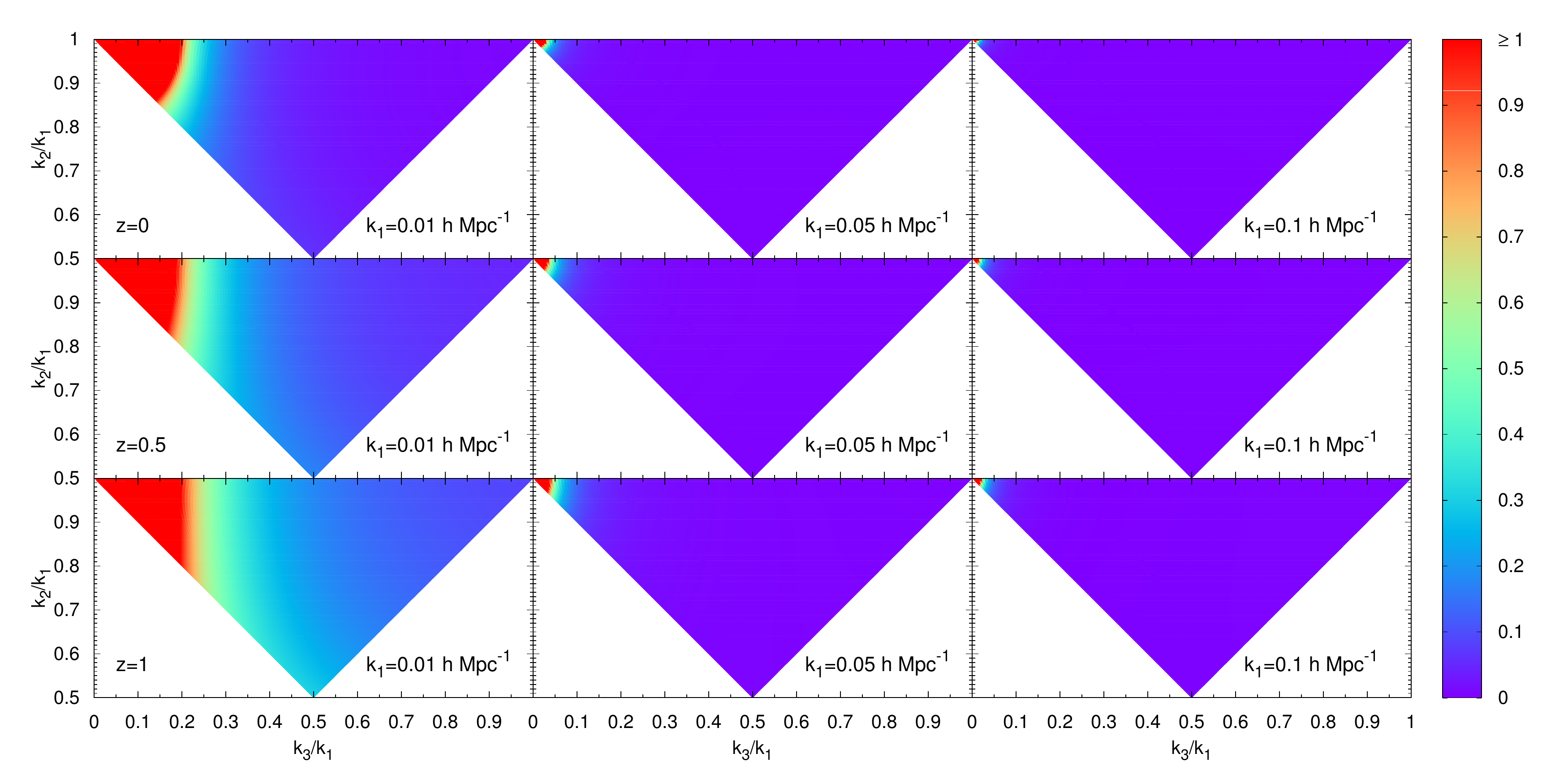}
\caption{The plots show the absolute value of the relative difference between the galaxy bispectrum monopole with PNG and the one without, for objects with mass $M = 10^{13} h^{-1} M_{\odot}$ and $\fnl =10$. The colour maps display the amplitude of the signal as a function of $k_2/k_1$ and $k_3/k_1$, under the condition $k_3 \leq k_2 \leq k_1$. Differences above $100\%$ are saturated to the same red colour of the palette.}
\label{fig:diff}
\end{figure}

In order to provide an illustration of the role played by PNG in redshift space, we plot in \cref{fig:diff} the absolute value of the relative difference between the non-Gaussian and the Gaussian monopole, 
\begin{equation}
 \text{Diff}(\k_1,\k_2,\k_3) = \, \Biggr\vert \frac{B_{ggg}^{s\,(0,0)} - B_{ggg}^{s\,G\,(0,0)}}{B_{ggg}^{s\,G\,(0,0)}} \Biggr\vert \,,
\end{equation}
assuming $\fnl = 10$ and objects with mass $M=10^{13} h^{-1} M_{\odot}$. The plots are based on the graphical representation of \cite{Jeong:2009}, i.e.~the amplitude of the signal is presented in a colour map as a function of $k_2/k_1$ and $k_3/k_1$, under the condition $k_3 \leq k_2 \leq k_1$, which avoids multiple visualizations of the same triangle/configuration.

The primordial non-Gaussian signal clearly peaks in the squeezed limit (top left corners) of the galaxy monopole, mainly on large scales and/or high redshift. On the
other hand, interestingly,
 a non-negligible signal propagates also into other configurations, with decreasing amplitude as it approaches the equilateral configuration (top right corners). 
  We interpret this effect, at least partly, as due to the new contributions FoGnG discussed above; primordial non-Gaussianity in redshift space induces distortions
   that can affect large scale measurements.    
 Failing to include all the non-Gaussian effects together with RSD would result into biased measurements of $\fnl$.

\section{Fisher Analysis}\label{sec:fisher}
As the recent measurement of Gil-Mar{\'{\i}}n {\em et al} \cite{Gil-Marin:2014sta,Gil-Marin:2014baa} shows, the bispectrum is a valuable tool for cosmology. In particular, the larger amount of available configurations in a wavelength range between $k_\text{min}$ and $k_\text{max}$ compared to the power spectrum may well shrink the observational bounds on $\fnl$ even further, as many works show \cite{Scoccimarro:2003wn,Sefusatti:2007ih,Baldauf:2010vn,Sefusatti:2012}, potentially below the \emph{Planck} constraint \cite{Baldauf:2016sjb}. The monopole derived in the previous section naturally extends the model considered by Gil-Mar{\'{\i}}n {\em et al} \cite{Gil-Marin:2014pva} to the case of local-type PNG and can be used to measure $\fnl$.

However, it may not be the only way to improve our constraints on inflation: the multi-tracer technique is a promising tool~\cite{Seljak:2008xr}. Even the novel position-dependent power spectrum could be an interesting alternative, although a detailed analysis in the case of PNG is still missing (see the discussion in \cref{app:pdps}). As both involve measurements of the power spectrum, they require less efforts than a full bispectrum analysis.

In this section we present forecasts of the accuracy in determining $\fnl$ -- a quantity that we call $\sigma_{\fnl}$ -- based on a Fisher analysis for the bispectrum.  Our aim is to give an illustration of the best possible improvement one could get with respect to power spectrum forecasts, when a single tracer is considered \cite{Font-Ribera:2013rwa,Zhao:2015gua}. On the other hand, important effects (like the covariance between different triangles) will be neglected and the quoted results are by no means intended to be fully realistic; rather, they are meant to motivate future work. For this reason we will consider only the bispectrum in real space, which is computationally easier to handle than the redshift space result.
Although forecasts based on the redshift-space bispectrum can have some small quantitative differences respect to our results, we expect the following qualitative discussion to hold anyway\footnote{Technically, studies on the cumulative signal-to-noise, i.e.~summed over all the configurations, show the critical dependence of the halo bispectrum signal on some kind of triangle configurations and the maximum wavenumber, $k_{\text{max}}$, considered \cite{Baldauf:2010vn,Sefusatti:2012}. At large scales ($k_{\text{max}} < 0.05 h$Mpc$^{-1}$), the signal is strongly suppressed because only few configurations are available, and with a large variance. By increasing $k_{\text{max}}$, the number of triangles considerably grows ($N_{\text{Tr}} \sim k_{\text{max}}^3$) and, consequently, the signal. As \cref{fig:diff} suggests, a large fraction of it is in squeezed configurations. Among these, the FoGnG term can play a role on those that have their smallest $k$ on sufficiently large scales. However, one should bear in mind that these large-scale, squeezed triangles are highly correlated, i.e.~the covariance cannot be neglected (see also the discussion in \cref{sec:method}). Thus, we expect that forecasts based on the redshift-space bispectrum will have small differences respect to our results, but it is clear that failing to include RSD in the bispectrum model could bias a $\fnl$ measurement at the level of accuracy that is now required.}.

The Fisher formalism is a tool for setting a lower limit on the statistical uncertainties that future surveys will have in the measurements of cosmological parameters of interest (see \cite{Verde:2007wf,2009arXiv0906.0664H,2010LNP...800..147V} for an introduction). The Fisher matrix information is defined as
\begin{equation}
 F_{\alpha \beta} \equiv \,- \biggr \langle \frac{\partial^2 \ln L (\x;p) }{\partial p_\alpha \partial p_\beta} \biggr \rangle \,,
\end{equation}
where $L (\x;p)$ is the likelihood function, i.e the probability of the data $\x$ given the parameters $p$, and $p_\alpha$ is the $\alpha$-th unknown parameter. If all the parameters are fixed except one (say $p_\alpha$), then the lower limit on the $1\sigma$ error bar in the $p_\alpha$ measurement is $\sigma_{p_\alpha} = 1/\sqrt{F_{\alpha \alpha}}$. Otherwise, if we marginalise over the parameters, the lower bound becomes $\sigma_{p_\alpha} = \sqrt{F_{\alpha \alpha}^{-1}}$.

The Fisher matrix for the bispectrum is \cite{Scoccimarro:2003wn,Baldauf:2016sjb}
\begin{equation}
 F_{\alpha \beta} \equiv \sum_{z} \sum_{k_1, k_2, k_3 \geq k_{\text{min}}}^{k_{\text{max}}} \frac{1}{\Delta B^2 (k_1, k_2, k_3)}\frac{\partial B (k_1, k_2, k_3)}{\partial p_\alpha} \frac{\partial B (k_1, k_2, k_3)}{\partial p_\beta}
\end{equation}
where $B$ and $\Delta B^2$ are the bispectrum and covariance estimator respectively and we assume the minimum value of $k$ to be fixed by the survey volume $V$, $k_{\text{min}} = 2 \pi / V^{1/3}$, while the maximum is $k_{\text{max}} = 0.1 D(0)/D(z)$ -- a reasonable limit for the validity of the non-linear analytic model \cite{Zhao:2015gua}.
 
\subsection{Methodology}\label{sec:method}
To compute the Fisher matrix we need to define $B$, $\Delta B^2$ and the set of unknown parameters $p$; our assumptions are described below.

We assume the bispectrum model of \cref{eq:bggg}, while the covariance for a survey of volume $V$ is given by \cite{Scoccimarro:1997st}
\begin{equation}
 (\Delta B)^2 = s_{123} \frac{V_f}{V_{123}} \left(P_{gg}(k_1) + \frac{1}{\bar{n}} \right) \left(P_{gg}(k_2) + \frac{1}{\bar{n}}  \right) \left(P_{gg}(k_3) + \frac{1}{\bar{n}}  \right) \,,
\end{equation}
where the volume of the fundamental cell is $V_f = (2 \pi)^3 / V$ and $V_{123} \approx 8 \pi^2 k_1 k_2 k_3 \delta k^3$, with $\delta k$ being the bin size. $\bar{n}$ is the number density of objects accounting for the shot noise, while $s_{123}$ is the symmetry factor, respectively $s_{123}=6,2,1$ for equilateral, isosceles and general configurations. In the noise estimator, the power spectrum is approximated by the leading contribution $P_{gg} (k) = \left(b_{10} + b_{01}/\alpha(k) \right) ^2P(k)$. 

 For simplicity and as in other studies \cite{Scoccimarro:2003wn,Sefusatti:2007ih,Sefusatti:2012,Baldauf:2010vn,Liguori10}, we neglect the covariance between different configurations of the bispectrum which is expected to be non-negligible for triangles sharing one or two sides, in particular on large scales \cite{Sefusatti:2006pa}. The induced covariance arising from survey selection effects (i.e.~complicated survey geometry and mask) is ignored as well. 

As explained in \cite{Sefusatti:2007ih}, the results of \cite{Scoccimarro:2003wn} suggest that ignoring covariance can over-estimate the constraining power of a given sample by a factor of two for $k < 0.1h$Mpc$^{-1}$ and up to a factor of eight for $k < 0.3h$Mpc$^{-1}$ at redshift zero. At higher redshift the contribution from a connected $6$-point function generated by non-linear gravitational evolution is expected to be less important and thus the (theoretical) covarince reduced. The largest $k$ values used in our forecasts lie between 0.15$h$Mpc$^{-1}$ and 0.25$h$Mpc$^{-1}$ in the redshift range $0<z<2.2$; suggesting covariance could make our forecasts optimistic by a factor of $\sim 5$. However, we will show that the constraining power in current and future surveys would provide competitive $\fnl$ constraints even if the covariance degrades our idealised forecasts by a factor of $5$, although a more realistic analysis is needed to fully explore this.

We assume all the cosmological parameters to be fixed to \emph{Planck}'s central values \cite{Ade:2015xua}, except for the linear and non-linear bias and $\fnl$, thus $p = \{ b_{10}, b_{20}, \fnl \}$. The fiducial model that maximizes the likelihood is assumed to be $p = \{ b_{10}^{\text{fid}}, b_{20}^{\text{fid}}, \fnl^{\text{fid}} = 0 \}$, where $b_{10}^{\text{fid}}$ is calibrated against real data (either from the particular survey or characteristic of the type of galaxy expected to be observed by the particular survey) and will be quoted in the following paragraphs, depending on the survey and the tracer. $\fnl^{\text{fid}}$ is assumed to be vanishing, this being compatible with current data: the final results will give an idea of the significance level at which a non-null primordial signal can be detected. 

Since the non-linear bias is not a well constrained parameter, the choice of the fiducial value is very important. We take it to be the analytic prediction $b_{20}^{\text{fid}} = b_{20}(\nu)$, based on \cref{eq:eul_bias,eq:b10,eq:b20}, where the variable $\nu$ is estimated under the assumption $b_{10}(\nu) =  b_{10}^{\text{fid}}$. We will then show how much $\sigma_{\fnl}$ is affected by the choice of $b_{20}^{\text{fid}}$ by allowing for a $\pm 1$ range around this value. Therefore, the results are presented in \cref{tab:others_fnl} in the following form:
\[
 {\sigma_{\fnl, b_{20}^{\text{fid}}}}^{\left( \sigma_{\fnl, b_{20}^{\text{fid}}+1} \right)}_{\left( \sigma_{\fnl, b_{20}^{\text{fid}}-1} \right)} ~.
\]
In our analysis we consider four redshift surveys: BOSS, eBOSS, DESI and Euclid, briefly presented below. The data used for each of them can be found in the tables in \cref{app:data}.

\subsubsection*{BOSS}
SDSS-III's Baryon Oscillation Spectroscopic Survey \cite{BOSS} is a galaxy redshift survey, which finished observations in $2014$. It mapped the spatial distribution of about $1.5$ million luminous red galaxies (LRGs)\footnote{Strictly speaking, BOSS also contains a sample of luminous galaxies with more star formation and greater disk morphology than typical LRGs.}, covering $10,000 \degs$ in the redshift range $0 < z < 0.8$, with the primary goal of detecting the characteristic scale imprinted by sound waves in the early universe, i.e.~the Baryon Acoustic Oscillations (BAO). Also, about $160,000$ quasars (QSOs) were observed in the redshift range $2.2 < z <3$, so that correlations can be measured in the Lyman-$\alpha$ forest, which we will not consider here. \Cref{tab:BOSS} in \cref{app:data} shows the basic numbers for BOSS, with the linear bias assumed to be $b_{10}^{\text{LRG}} = 1.7/D(z)$ (see \cite{Font-Ribera:2013rwa} and references therein).

\subsubsection*{eBOSS}
The extended Baryon Oscillation Spectroscopic Survey \cite{Dawson:2015wdb} is part of the SDSS-IV project and started observations in $2014$. It will extend the BAO measurements to $0.6 < z < 2.2$ by observing LRGs, Emission Line Galaxies (ELGs) and QSOs.

The eBOSS numbers we use match those presented in \cite{Zhao:2015gua}. LRGs will be observed in the redshift range $0.6 < z < 1$ over $7,000 \degs$, with a linear bias assumed to be $b_{10}^{\text{LRG}} = 1.7/D(z)$, while QSOs will fall in the range $0.6 < z < 2.2$ over $7,500 \degs$, with bias $b_{10}^{\text{QSO}} = 0.53 + 0.29 (1 + z)^2$. The ELG target selection definitions have not been finalised, but each of the three proposals considered in \cite{Zhao:2015gua} result in samples that have a significant overlap in volume with the LRG sample. We have tested each potential ELG sample and found that even if they are treated independently, they do not add substantial constraining power. Therefore, we omit them from the forecast constraints we present. \Cref{tab:eBOSS_LRGs,tab:eBOSS_QSO,tab:eBOSS_ELGs} show the basic numbers for eBOSS LRGs, QSOs and ELGs respectively. Refer to \cite{Zhao:2015gua} and references therein for further details.

\subsubsection*{DESI}
Dark Energy Spectroscopic Instrument \cite{DESI} is a redshift survey with the primary target of measuring the effect of dark energy on the expansion of the Universe. It is expected to run between $2018$ and $2022$ and will map the universe from low to high redshift over $14,000 \degs$, measuring the optical spectra for tens of million objects, including LRGs, ELGs and QSOs. 

The LRGs will fall in the redshift range $0.1 < z < 1.1$ with a linear bias assumed to be $b_{10}^{\text{LRG}} = 1.7/D(z)$, ELGs in $0.1 < z < 1.8$ with $b_{10}^{\text{ELG}} = 0.84/ D(z)$ and QSOs will be considered in redshift range $0.1 < z < 1.9$, with bias $b_{10}^{\text{QSO}} = 1.2/ D(z)$. \Cref{tab:DESI} shows the basic numbers for DESI (see \cite{Font-Ribera:2013rwa} and references therein).

\subsubsection*{Euclid}
Euclid \cite{EUCLID} is a space mission developed to study the imprints of dark energy and gravity. The expansion rate of the Universe and the growth of structures will be tracked by using two complementary observables: weak gravitational lensing and galaxy clustering. Its launch is planned for $2020$.

We focus on the redshift survey part of the mission, which is expected to detect about $50$ million galaxies in the redshift range $0.6 < z < 2.1$, over $15,000 \degs$. The fiducial value for the bias is assumed to be $b_{10} = 0.76/D(z)$. \Cref{tab:Euclid} shows the basic numbers for Euclid (see \cite{Font-Ribera:2013rwa} and references therein).

\subsection{Results}\label{sec:reults}
Given the assumptions listed above, \cref{tab:others_fnl} shows the lower limit on $\sigma_{\fnl}$ that could be expected from the bispectrum of BOSS, eBOSS, DESI and Euclid. We also present the forecast results for the power spectrum, in order to provide a comparison. The results combine all the tracers available for each survey (LRGs, ELGs, QSOs), which are treated as independent, except that we omit any ELG sample for eBOSS, as previously noted.

If we focus first on the results labelled `bias float' (marginalising over bias) of the bispectrum set of columns, our analysis suggests that BOSS and eBOSS will both be able to reach $\sigma_{\fnl} \simeq 1$. eBOSS appears to be penalised compared to BOSS because of the lower number densities. Interestingly, DESI and Euclid may give $\sigma_{\fnl} < 1$, regardless of the chosen fiducial value for $b_{20}$, within $\pm 1$ range. The DESI result is more stringent than the Euclid one because DESI is assumed to observe more biased objects; however, combining the BOSS and Euclid data tighten the constraint towards the DESI result. Clearly, the $\fnl$ measurements improve by fixing the linear and non-linear bias: the results on the last column (bias fixed) of \cref{tab:others_fnl} indicate an improvement factor between $1.4$ and $2$ on $\sigma_{\fnl}$.

A comparison between $\sigma_{\fnl}$ expected from the bispectrum and those from the power spectrum of a single tracer (first two columns of \cref{tab:others_fnl}, but see also \cite{Font-Ribera:2013rwa,Zhao:2015gua}) seems to indicate about an order of magnitude improvement. Even allowing for a factor of five dilution in constraining power potentially caused by covariance between triangle configurations, our forecasts remain impressive. Indeed, we forecast that current BOSS data should allow $\fnl$ constraints competitive with those obtained from \emph{Planck} \cite{Ade:2015ava}.

\begin{table}
\centering
\caption{Forecasts for $\sigma_{\fnl}$ from the bispectrum of BOSS, eBOSS, DESI and Euclid, assuming the fiducial values $p = \{ b_{10}^{\text{fid}}, b_{20}^{\text{fid}}, \fnl^{\text{fid}} = 0 \}$, as described in \cref{sec:method}. Forecasts from the power spectrum are obtained considering only the tree-level, with the fiducial model $p = \{ b_{10}^{\text{fid}}, \fnl^{\text{fid}} = 0 \}$. The results with marginalisation over the bias factors are shown on the left columns (bias float), while those without on the right (bias fixed). The numbers inside the parenthesis in the superscripts are the predictions for $\sigma_{\fnl}$ considering the fiducial value for the non-linear bias to be $b_{20}^{\text{fid}}+1$, while those in the subscripts assume $b_{20}^{\text{fid}}-1$.}    
\label{tab:others_fnl}
\begin{tabular}{ccccc}
\hline\hline
		      &  \multicolumn{2}{c}{Power Spectrum}	&   \multicolumn{2}{c}{Bispectrum} \\
Sample		      & $\sigma_{\fnl}$ & $\sigma_{\fnl}$ 	& $\sigma_{\fnl}$ 	  		  & $\sigma_{\fnl}$  			\\
		      & bias float	    & bias fixed	& bias float		  		  & bias fixed 				\\  
\hline
BOSS 	  	      &  $21.30$	    & $13.28$	        & $ 1.04^{(0.65)}_{(2.47)} $		  &  $ 0.57^{(0.35)}_{(1.48)} $        \\  
eBOSS          &  $14.21$        & $11.12$          & $ 1.18^{(0.82)}_{(2.02)} $          &  $ 0.70^{(0.48)}_{(1.29)} $        
\\
Euclid 	  	  &  $6.00$	    &  $4.71$		& $ 0.45^{(0.18)}_{(0.71)} $		  &  $ 0.32^{(0.12)}_{(0.35)} $        \\   
DESI  		  &  $5.43$	    &  $4.37$		& $ 0.31^{(0.17)}_{(0.48)} $		  &  $ 0.21^{(0.12)}_{(0.37)} $        \\   
\hline
BOSS + Euclid     &  $5.64$	    &  $4.44$		&$ 0.39^{(0.17)}_{(0.59)} $	   	  &  $ 0.28^{(0.11)}_{(0.34)} $        \\  
\hline\hline  
\end{tabular}
\end{table} 

\section{Conclusions}\label{sec:discussion}
The target sensitivity of ${\fnl} \sim 1$ sets a new challenge in the search for primordial non-Gaussianity.
While future CMB experiments may not be able to achieve this goal, large-scale structure observations hold the promise to reach this level of sensitivity, by exploiting the characteristic scale-dependence introduced by local-type models in the bias relation between collapsed objects and the density field, and the very large amount of data available with future redshift surveys.

In this work, we have studied the sensitivity of galaxy bispectrum measurements to $\fnl$, tackling the problem from two separate directions. We first addressed the problem of modelling redshift space distortions in the tree-level galaxy bispectrum with primordial non-Gaussianity of local-type. We
examined how redshift space distortions
 can affect large-scale measurements, and therefore potentially lead to a biased measurement of $\fnl$
  if not properly described. In particular, we identified new contributions to the galaxy
  bispectrum, which physically correspond to large-scale amplifications -- induced
  by 
  primordial non-Gaussianity -- 
   of redshift space distortion effects. 
    Moreover, 
    we proposed an analytic prediction for the monopole which can be used to fit against data, in the large scale regimes where the non-linear part of FoG can be neglected. 
    We analysed the 
    physical consequences of our findings, providing a graphical method for comparing our results for bispectra  
   with the case in which primordial non-Gaussianity is not included. 
    
    We then performed 
     idealised forecasts of 
     $\sigma_{\fnl}$, the accuracy
      of the determination of local $\fnl$, that could be obtained from 
       measurements of
      the galaxy bispectrum using data from 
      surveys like BOSS, eBOSS, DESI and Euclid. 
Our findings suggest that the bispectrum of galaxies in current and future surveys will provide competitive $\fnl$ constraints even if the covariance between triangle configurations degrades our idealised forecasts by a factor of $5$. In particular, current BOSS data should allow for \emph{Planck}-like constraints on $\fnl$, while future surveys like Euclid and DESI will contain the statistical power to shrink the bound by an additional factor of three.

We leave as a challenge for future work to obtain improved predictions for $\sigma_{\fnl}$ fully accounting for the covariance: this will be necessary if we are to completely understand the power of bispectrum measurements to constrain $\fnl$ compared to alternative approaches, such as the multi-tracer technique or the position-dependent power spectrum.

\acknowledgments
MT wishes to thank H\'ector Gil-Mar\'in, Yuting Wang, Benedict Kalus, Davide Bianchi and Gong-Bo Zhao for many helpful discussions on the topics of redshift space distortions and Fisher analysis. He also acknowledges Marco Crisostomi, for help with \emph{Mathematica} and sharing his enthusiasm, and Donough Regan for providing useful comments. 
GT  is supported by STFC grant ST/N001435/1.
DW is supported by STFC grants ST/K00090X/1 and ST/L005573/1.

\bibliographystyle{JHEP}
\bibliography{biblio}

\appendix
\section{Halo mass function and Lagrangian bias}\label{app:massfunction}
Through this paper we assume a mass function,
\begin{equation}
  n_g (\MA,z) = f(\nu) \frac{\rho_m}{\MA} \biggr\vert \frac{d \ln \sigma}{d \MA} \biggr\vert \,,
\end{equation}
with $f = f_\ST \, ( f_\LV / f_\PS)$. The Press-Schechter mass function $f_\PS$ \cite{1974ApJ...187..425P} for Gaussian initial conditions and spherical collapse is 
\begin{equation}\label{eq:ps}
  f_\PS (\nu) = \sqrt{\frac{2}{\pi}} \nu e^{-\frac{\nu^2}{2}} \,,
\end{equation}
while, allowing for ellipsoidal collapse, the Sheth-Tormen mass function $f_\ST$ \cite{Sheth:1999,Sheth:2001,Sheth:2002} is obtained:
\begin{equation}\label{eq:st}
  f_\ST (\nu) = A(p) \sqrt{\frac{2 \gamma}{\pi}} \left[ 1 + \left( \gamma \nu^2 \right)^{-p}\right] \nu e^{-\gamma \, \frac{\nu^2}{2}} \,.
\end{equation}
By fitting against simulations, one finds the parameters $\gamma = 0.707$ and $p = 0.3$. Then, requiring all the mass to be collapsed into halos gives $A(p) = 0.322$.

By approximating a weakly non-Gaussian initial state with an Edgeworth expansion, the Lo Verde {\em et al} mass function $f_\LV$ \cite{LoVerde:2007ri} is found
\begin{equation}\label{eq:lv}
  f_\LV(\nu,\MA) = f_\PS(\nu) \left[ 1 + \frac{1}{6}\left(\kt(\MA) H_3(\nu) - \frac{d\kt(\MA)/d\MA}{d\ln\sigma^{-1}/d\MA} \frac{H_2(\nu)}{\nu} \right) \right] \,,
\end{equation}
where the function $H_n$ is the $n$-th Hermite polynomial and the $3$rd cumulant $\kt(\MA)$ is defined as $\kt(\MA) = \langle \delta_\lin^3 \rangle / \sigma^3$ with
\begin{equation}
\begin{split}
 \langle \delta_\lin^3 \rangle = \int \dqc \int \dqcp \int \dqcpp W_\MA(p) \alpha(p,z) W_\MA(p') \alpha(p',z) W_\MA(p'') \alpha(p'',z) \times \\ \times \langle \Phi_\ini(\p) \Phi_\ini(\p') \Phi_\ini(\p'') \rangle \,.
\end{split}
\end{equation}
A convenient fitting function for $\kt(\MA)$ is given in \cref{eq:kt}. Also, for primordial non-Gaussianity of the form of \cref{eq:png}, the variance gets a correction proportional to $\fnl^2$, that can be written as
\begin{equation}\label{eq:kd2}
  \begin{split}
      \sigma^2 = \langle \delta_\lin^2 \rangle & = \int \dqc \int \dqcp W_\MA(p) \alpha(p,z) W_\MA(p') \alpha(p',z) \langle \Phi_\ini(\p) \Phi_\ini(\p') \rangle \\
					    & \approx \sigma_G^2 \left( 1 + \kd(\MA) \right) \,.
  \end{split}
\end{equation}
Since $\kd$ gives a negligible correction to the Gaussian variance $\sigma_G = \langle \delta_\G^2 \rangle$ for any realistic value of $\fnl$ \cite{2011JCAP...08..003L}, we neglect it. Although the LV mass function is no longer universal, as it formally depends not only on $\nu$ but also on the mass $\MA$, the extra dependence on $\MA$ through \cref{eq:kd2,eq:kt} is weak and we practically treat $f_\LV$ as universal.

As explained in \cref{sec:bivariate}, local type PNG introduces in the mass function an additional dependence on the local effective variance $\sigma_l$, so that the halo/galaxy overdensity 
acquires a 
 bivariate form 
\begin{equation}
\begin{split}
 \delta_g^{\La}(\q)=\,& \beta_{10} \delta_{\lin,l} + \beta_{01} \left(\frac{\sigma_l}{\sigma}-1\right) +\\&+\frac{1}{2}\left[\beta_{20} (\delta_{\lin,l})^2 + \beta_{02} \left(\frac{\sigma_l}{\sigma}-1\right)^2 + 2 \beta_{11} \delta_{\lin,l} \left(\frac{\sigma_l}{\sigma}-1\right) \right] \,,
\end{split}
\end{equation}
with the bias coefficients defined as
\begin{equation}
 \label{eq:betaij}
 \beta_{ij} \equiv \left[\frac{(\sigma_l)^j}{n_h}\frac{\partial^{i+j}n_g}{\left( \partial\delta_{\lin,l} \right)^i \left( \partial \sigma_l \right)^j} \right] \biggr\vert_{\delta_{\lin,l}=0, \sigma_l=\sigma} \, .
\end{equation}
By using the explicit form of \cref{eq:sigmal} for $\sigma_l$ and redefining the bias as
\begin{equation}\label{eq:newlagrangianbias}
  \begin{split}
    b_{10}^\La & = \beta_{10} \,, \\
    b_{01}^\La & = 2 \fnl \beta_{01}  \,,\\
    b_{20}^\La & = \frac{\beta_{20}}{2}  \,,\\
    b_{11}^\La & = 2 \fnl \beta_{11}  \,,\\
    b_{02}^\La & = 2 \fnl^2 \beta_{02} \,,
  \end{split}
\end{equation}
that we call \emph{Lagrangian bias coefficients}, one can then easily obtain the bivariate model in the form quoted in \cref{eq:bivariate}. By using the mass function of \cref{eq:stlv} in the definition of \cref{eq:betaij}, the linear and non-linear Lagrangian bias coefficients read
\begin{align}
      b_{10}^\La & = \frac{\gamma \nu^2 -1}{\delta_c} + \frac{2p}{1+(\gamma \nu^2)^p}\frac{1}{\delta_c} - \kt \frac{\nu^3 - \nu}{2\delta_c} + \dkt \frac{\nu + \nu^{-1}}{6 \delta_c} \label{eq:b10} \\
      b_{20}^\La & = \gamma \nu^2 \frac{\gamma \nu^2 - 3}{2 \delta_c^2} + \frac{p}{1+(\gamma \nu^2)^p} \frac{2\gamma\nu^2 + 2p -1}{\delta_c^2} - \frac{\kt}{2} \left[ \frac{\gamma \nu^5 -(\gamma+2)\nu^3 + \nu}{\delta_c^2} + \frac{2p}{1+(\gamma \nu^2)^p} \frac{\nu^3 - \nu}{\delta_c^2}\right] + \nonumber \\ & \quad + \frac{1}{2} \dkt \left[ \frac{\gamma\nu^3 + (\gamma -1) \nu}{3\delta_c^2} + \frac{2p}{1+(\gamma \nu^2)^p} \frac{\nu - \nu^{-1}}{3\delta_c^2} \right] \label{eq:b20} \,,
\end{align}
while all the non-Gaussian bias factors are built from a combination of these:
\begin{equation}
  \begin{split}
    b_{01}^\La & = 2 \fnl \delta_c b_{10}^\La  \,,\\
    b_{11}^\La & = 2 \fnl (\delta_c b_{20}^\La - b_{10}^\La)  \,,\\
    b_{02}^\La & = 4 \fnl^2 \delta_c (\delta_c b_{20}^\La - 2b_{10}^\La)  \,.  
  \end{split}
\end{equation}

\section{Position-dependent power spectrum}\label{app:pdps}
Although the bispectrum contains more information than power spectrum, it is more challenging to measure and, indeed, only few measurements have been reported so far \cite{Verde:2001sf,Scoccimarro:2003wn,Jing:2003nb,Gaztanaga:2005an,FMarin:2011,McBride:2011,Marin:2013bbb,Gil-Marin:2014sta,Gil-Marin:2014baa}. To overcome this issue, a new observable has been proposed in \cite{Chiang:2014oga}, which measures an integral of the squeezed configuration of the bispectrum. In \cite{Chiang:2015eza}, this has been applied to measure the non-linear bias $b_{20}$ from the BOSS data release $10$. 

This new observable, called \emph{position-dependent power spectrum}, correlates the power spectrum in a subvolume of the survey volume to the mean overdensity of the subvolume itself: basically it measure the response of the spectra of short density modes to a large-scale fluctuation. We briefly describe below how the position-dependent power spectrum is built,
 referring  the reader to \cite{Chiang:2014oga,Chiang:2015eza} for further details.

Given a density field $\delta(\x)$ in a cubic survey volume $V$ with length side $L_B$, suppose to split it into $N$ subvolumes, with side $L = L_B / N$. If we now focus on the subvolume centred at $\x_L$, we can measure the local mean overdensity as
\begin{equation}
 \bar\delta(\x_L) = \frac{1}{V_L}\int d^3x~\delta(\x)W_L(\x-\x_L) ~,
\label{eq:deltabar}
\end{equation}
where the volume of the subvolume is $V_L = L^3$ and the window function is assumed to be
\begin{equation}
W_L(\x)=\prod_{i=1}^3\:\theta(x_i), \quad
\theta(x_i) = \left\{
\begin{array}{cc}
1, & |x_i|\le L/2, \\
0,  & \mbox{otherwise}~.
\end{array}\right.
\label{eq:WL}
\end{equation}
The Fourier transform of the window is $W_L(\k)=L^3\prod_{i=1}^3 j_{0}(k_iL/2)$, where the $0$-th spherical Bessel function is $j_{0} (x)=\sin(x)/x$. The position-dependent power spectrum is defined as
\begin{equation}
P(\k,\x_L)\equiv \frac1{V_L}|\delta(\k,\x_L)|^2\,,
\end{equation}
where $\delta(\k,\x_L)\equiv \int_{V_L}d^3x~\delta(\x)e^{-i\x\cdot\vk}$ is the Fourier transformation of the density field with integral ranging over the subvolume centred at $\x_L$.

If we now correlate the mean overdensity to the position-dependent power spectrum in the corresponding subvolume, it can be shown that
\begin{eqnarray}
 \langle P(\k,\x_L)\bar\delta(\x_L)\rangle&=&\frac{1}{V_L^2}\int\frac{d^3p_1}{(2\pi)^3}\int\frac{d^3p_3}{(2\pi)^3}~
 B(\k-\p_1,-\k+\p_1+\p_3,-\p_3) \times\\
\nonumber 
& &\hspace{3.3cm}\times W_L(\p_1)W_L(-\p_1-\p_3)W_L(\p_3) \\
&\equiv& iB(\k)~,
\label{eq:corr_localpk_deltabar}
\end{eqnarray}
where $iB(\k)$ is called the \emph{integrated bispectrum} and $B(\k_1,\k_2,\k_3)$ can be the matter bispectrum, the galaxy bispectrum or cross-correlations between matter and galaxies. An angular average over $iB(\k)$ removes the remaining $\hat{\k}$-dependence due to the choice of a cubic window function and one finally gets
\begin{eqnarray}
 iB(k)&\equiv&\int\frac{d^2\Omega_{\hat\k}}{4\pi}~iB(\k)
 =\frac{1}{V_L^2}\int\frac{d^2\Omega_{\hat\k}}{4\pi}\int\frac{d^3p_1}{(2\pi)^3}
 \int\frac{d^3p_3}{(2\pi)^3}~B(\k-\p_1,-\k+\p_1+\p_3,-\p_3) \times \nonumber\\
& &\hspace{4.8cm}\times  W_L(\p_1)W_L(-\p_1-\p_3)W_L(\p_3) ~. 
\label{eq:iB_angular_average}
\end{eqnarray}
From the behaviour of the $0$-th spherical Bessel function, the dominant contribution to the integrated bispectrum comes from wavenumbers $\k$ that are larger than $1/L$, i.e. from the squeezed configuration of the bispectrum  $B(\k-\p_1,-\k+\p_1+\p_3,-\p_3) \to B(\k,-\k,-\p_3)$ with $p_1\ll k$ and $p_3\ll k$.

If we consider the tree-level matter bispectrum with Gaussian initial conditions ($\fnl=0$),
\begin{equation}
B_{mmm}^G(\k_1,\k_2,\k_3)=2[P(k_1) P(k_2) F_2(\k_1,\k_2) + 2 \:\rm perm], 
\label{eq:Bspt}
\end{equation}
it can be shown that the integrated bispectrum is \cite{Chiang:2014oga}
\begin{eqnarray}
 iB_{mmm}^G(k) &\stackrel{k L \to \infty}{=}&
 \left[\frac{68}{21}-\frac{1}{3}\frac{d\ln k^3 P(k)}{d\ln k}\right]P(k)\sigma_L^2 ~,
\label{eq:int_bi_approx}
\end{eqnarray}
where $\sigma_L^2$ is the variance of the density field on the subvolume scale,
\begin{equation}
\sigma_L^2 \equiv \frac{1}{V_L^2}  \int\frac{d^3p_3}{(2\pi)^3}
W_L^2(\p_3)  P(p_3)~.
\label{eq:sigmaL}
\end{equation}
By including local type PNG, the tree-level matter bispectrum reads
\begin{equation}
B_{mmm}(\k_1,\k_2,\k_3) = 2 \left[\mathcal{F}_2(\k_1,\k_2) +\fnl\frac{\alpha(k_3)}{\alpha(k_1)\alpha(k_2)}\right]P(k_1)P(k_2) + 2\cyc 
\end{equation}
and the linear response of the small-scale matter power spectrum to large-scale density perturbation is now
\begin{equation}
 iB_{mmm}(k) \stackrel{k L \to \infty}{=} \left[\frac{68}{21}-\frac{1}{3}\frac{d\ln k^3 P(k)}{d\ln k}\right]P(k)\sigma_L^2 + 4 \fnl \sigPNGU P(k) + 2 \fnl \sigA \frac{P^2(k)}{\alpha^2(k)} \label{eq:pdibm}\,,
\end{equation}
where we have introduced the new quantities
\begin{align}
 \sigma^2_{\text{nG},i} \equiv & \, \frac{1}{V_L^2}  \int\frac{d^3p_3}{(2\pi)^3} W_L^2(\p_3)  \frac{P(p_3)}{\alpha^i(p_3)} ~,\\
 \sigA \equiv & \, \frac{1}{V_L^2}  \int\frac{d^3p_3}{(2\pi)^3} W_L^2(\p_3)  \alpha(p_3) ~.
\end{align}

If we now consider the galaxy bispectrum of \cref{eq:bggg}, the integrated bispectrum is
\begin{align}
 iB_{ggg}(k) \stackrel{k L \to \infty}{=} & P(k)\sigma_L^2 \Biggr\{ \biggr[ b_{10}^3 \left(\frac{68}{21}-\frac{1}{3}\frac{d\ln k^3 P(k)}{d\ln k}\right) + 4 b_{10}^2 b_{20} \biggr] + \nonumber \\ & \qquad +\frac{1}{\alpha(k)}\biggr[ b_{10}^2 b_{11} + 2 b_{10} b_{01} b_{20} + b_{10}^2 b_{01} \left(\frac{68}{21}-\frac{1}{3}\frac{d\ln k^3 P(k)}{d\ln k}\right) \biggr] + 2 \frac{b_{10} b_{01} b_{11}}{\alpha^2(k)} \Biggr\} + \nonumber \\
+ & P(k)\sigPNGU \Biggr\{ \biggr[ b_{10}^2 b_{01} \left(\frac{68}{21}-\frac{1}{3}\frac{d\ln k^3 P(k)}{d\ln k}\right) + 2 b_{10}^2 b_{11} + 2 b_{10} b_{01} b_{20} + 4 \fnl b_{10}^3  \biggr] + \nonumber \\ & \qquad \qquad + \frac{1}{\alpha(k)} \biggr[ b_{10}^2 b_{01} \left(\frac{68}{21}-\frac{1}{3}\frac{d\ln k^3 P(k)}{d\ln k}\right) +4 b_{10}^2 b_{02} + 2 b_{10} b_{01} b_{11} + 4 b_{01}^2 b_{20} \biggr] + \nonumber \\ & \qquad \qquad + \frac{2}{\alpha(k)^2} \biggr[ b_{01}^2 b_{11} + 2 b_{10} b_{01} b_{02} \biggr] \Biggr\} + \nonumber \\
+ & P(k)\sigPNGD \Biggr\{ 2 \left( b_{10} b_{01} b_{11} + 2 \fnl b_{10}^2 b_{01} \right) + \nonumber \\ & \qquad \qquad \qquad + \frac{2}{\alpha(k)} \left( b_{01}^2 b_{11} + 2 b_{10} b_{01} b_{02} + 2 \fnl b_{10} b_{01}^2 \right) + 4\frac{b_{01}^2 b_{02}}{\alpha^2(k)} \Biggr\} + \label{eq:pdib} \\
+ & P^2(k)\sigA \Biggr\{ \frac{2 \fnl}{\alpha(k)} \left[ b_{10}^3 +\frac{1}{\alpha^3(k)} \left(b_{10}^2 b_{01} + b_{10} b_{01}^2 \right) \right]\Biggr\} + \nonumber \\
+ & \frac{P^2(k)}{V_L} \Biggr\{ 2 \left( b_{10}^2 b_{20} -\frac{4}{21} b_{10}^2 b_{10}^\La \right) + \frac{2}{\alpha(k)} \left( b_{10}^2 b_{11} + b_{10} b_{01} b_{20} - \frac{8}{21} b_{10} b_{01} b_{10}^\La + b_{10}^2 b_{01} \right)  + \nonumber \\ & \qquad + \frac{2}{\alpha^2(k)} \left(b_{10}^2 b_{02} +2 b_{10} b_{01} b_{11} + b_{01}^2 b_{20} - \frac{4}{21} b_{01}^2 b_{10}^\La + 2 b_{10} b_{01}^2 + 2 b_{01}^3 \right) + \nonumber \\ & \qquad  + \frac{2}{\alpha^4(k)} \left( b_{01}^2 b_{02} + 2 b_{10} b_{01} b_{02} \right) \Biggr\} \nonumber \,.
\end{align}
\Cref{eq:pdibm,eq:pdib} are our prediction for $iB(k)$ coming from the matter and galaxy bispectrum with local-type non-Gaussianity, respectively. 
In \cite{Chiang:2015eza}, an analysis on $iB_{ggg}$ shows poor constraints on $\fnl$ compared to those from the power spectrum. However, since the scale-dependent bias due to local-type non-Gaussianity was ignored there, it would be interesting to see how much the constraint on $\fnl$ from the position-dependent power spectrum would improve when the result of \cref{eq:pdib} is used. We leave this for future work. 

\section{$\mathcal{D}$ factors}\label{app:dfactors}
The factors $\mathcal{D}(k_i,k_j,\cos\theta_{ij},y_{ij},\beta)$ introduced in \cref{sec:gmono} are defined as the integrals below
\begin{align}
 \mathcal{D}_{\text{SQ}1} & = \frac{1}{4\pi}\int_{-1}^{+1} d\mu_1 \int_{0}^{2\pi} d\phi \, 2\left(1+\beta \mu_i^2\right)\left(1+\beta \mu_j^2\right) \\
  \mathcal{D}_{\text{NLB}} & = \mathcal{D}_{\text{SQ}1} \\
  \mathcal{D}_{\text{SQ}2} & = \frac{1}{4\pi}\int_{-1}^{+1} d\mu_1 \int_{0}^{2\pi} d\phi \, 2\beta \mu_k^2\left(1+\beta \mu_i^2\right)\left(1+\beta \mu_j^2\right) \\
   \mathcal{D}_{\text{FOG}} & = \frac{1}{4\pi}\int_{-1}^{+1} d\mu_1 \int_{0}^{2\pi} d\phi \, \beta \mu_k k_k \left(1+\beta \mu_i^2\right)\left(1+\beta \mu_j^2\right) \left[\beta \mu_k k_k \frac{\mu_i}{k_i}\frac{\mu_j}{k_j} - \left(\frac{\mu_i}{k_i} + \frac{\mu_j}{k_j}\right) \right]\\
  \mathcal{D}_{\text{nG}2} & = \mathcal{D}_{\text{SQ}1} \\
    \mathcal{D}_{\text{FoGnG}} & = -\frac{1}{4\pi}\int_{-1}^{+1} d\mu_1 \int_{0}^{2\pi} d\phi \, \beta \mu_k k_k \left(1+\beta \mu_i^2\right)\left(1+\beta \mu_j^2\right) \left(\frac{\mu_i}{k_i \aj}+\frac{\mu_j}{k_j\ai}\right)\\
  \mathcal{D}_{\text{SQ}1}^{\text{nG}1} & = \frac{1}{4\pi}\int_{-1}^{+1} d\mu_1 \int_{0}^{2\pi} d\phi \, 2\left(\frac{1+\beta \mu_i^2}{\aj} + \frac{1+\beta \mu_j^2}{\ai} \right) \\
  \mathcal{D}_{\text{NLB}}^{\text{nG}1} & = \mathcal{D}_{\text{SQ}1}^{\text{nG}1} \\
  \mathcal{D}_{\text{SQ2}}^{\text{nG}1} & = \frac{1}{4\pi}\int_{-1}^{+1} d\mu_1 \int_{0}^{2\pi} d\phi \, 2\beta \mu_k^2 \left(\frac{1+\beta \mu_i^2}{\aj} + \frac{1+\beta \mu_j^2}{\ai} \right) \\
  \mathcal{D}_{\text{FOG}}^{\text{nG}1} & = \frac{1}{4\pi}\int_{-1}^{+1} d\mu_1 \int_{0}^{2\pi} d\phi \, \beta \mu_k k_k \left(\frac{1+\beta \mu_i^2}{\aj}+\frac{1+\beta \mu_j^2}{\ai}\right) \left[\beta \mu_k k_k \frac{\mu_i}{k_i}\frac{\mu_j}{k_j} - \left(\frac{\mu_i}{k_i} + \frac{\mu_j}{k_j}\right) \right]\\
   \mathcal{D}_{\text{nG}2}^{\text{nG}1} & = \mathcal{D}_{\text{SQ}1}^{\text{nG}1} \\
  \mathcal{D}_{\text{FoGnG}}^{\text{nG}1} & = -\frac{1}{4\pi}\int_{-1}^{+1} d\mu_1 \int_{0}^{2\pi} d\phi \, \beta \mu_k k_k \left(\frac{1+\beta \mu_i^2}{\aj}+\frac{1+\beta \mu_j^2}{\ai}\right) \left(\frac{\mu_i}{k_i \aj}+\frac{\mu_j}{k_j\ai}\right)\\
  \mathcal{D}_{\text{SQ2}}^{\text{nG}1^2} & = \frac{1}{4\pi}\int_{-1}^{+1} d\mu_1 \int_{0}^{2\pi} d\phi \, \frac{2\beta \mu_k^2}{\ai \aj} \\
  \mathcal{D}_{\text{FOG}}^{\text{nG}1^2} & = \frac{1}{4\pi}\int_{-1}^{+1} d\mu_1 \int_{0}^{2\pi} d\phi \, \beta \frac{\mu_k k_k}{\ai \aj} \left[\beta \mu_k k_k \frac{\mu_i}{k_i}\frac{\mu_j}{k_j} - \left(\frac{\mu_i}{k_i} + \frac{\mu_j}{k_j}\right) \right] \\
  \mathcal{D}_{\text{FoGnG}}^{\text{nG}1^2} & = -\frac{1}{4\pi}\int_{-1}^{+1} d\mu_1 \int_{0}^{2\pi} d\phi \, \beta \frac{\mu_k k_k}{\ai \aj} \left(\frac{\mu_i}{k_i \aj}+\frac{\mu_j}{k_j\ai}\right) \,,
\end{align}
yielding the following results
\begin{align}
 \mathcal{D}_{\text{SQ}1} = \,& \frac{2}{15} \left[15 + 10 \beta +\beta ^2 \left(2 x_{ij}^2+1\right)\right]\\
  \mathcal{D}_{\text{NLB}} = \,& \mathcal{D}_{\text{SQ}1} \\
  \mathcal{D}_{\text{SQ}2} = \,&\frac{2 \beta}{105 \left(2 x_{ij} y_{ij} + y_{ij}^2+1\right)} \biggr[ 12 \beta ^2  y_{ij} x_{ij}^3 + 2 \beta  x_{ij}^2  (6 \beta +7) \left( y_{ij}^2 + 1 \right)+ \nonumber \\& \qquad \qquad + 2 x_{ij} y_{ij} \left(9 \beta ^2+42 \beta +35\right)  +\left(3 \beta ^2+28 \beta +35\right) \left( y_{ij}^2+1 \right) \biggr]  \\
   \mathcal{D}_{\text{FOG}} = \,& \frac{\beta}{315 y_{ij}}  \biggr[16 \beta ^3 y_{ij} x_{ij}^4  +4 \beta ^2 x_{ij}^3 (5 \beta +9) \left(y_{ij}^2+1\right)+24 \beta x_{ij}^2 y_{ij} \left(2 \beta ^2+9 \beta +7\right) + \nonumber \\& \qquad \qquad + 3x_{ij} \left(5 \beta ^3+33 \beta ^2+63 \beta +35\right)  \left(y_{ij}^2 + 1\right) + 6y_{ij}\left(\beta ^3+9 \beta ^2+35 \beta +35\right)\biggr] \\
  \mathcal{D}_{\text{nG}2} = \,&\mathcal{D}_{\text{SQ}1} \\
   \mathcal{D}_{\text{FoGnG}} = \,& \frac{\beta}{105 \ai \aj y_{ij}}  \biggr[6 \beta ^2 x_{ij}^3 \left(\au +\ad y_{ij}^2\right)+2 \beta  (6 \beta +7) x_{ij}^2 y_{ij} (\ai +\aj )+\nonumber \\ & + \left(9 \beta ^2+42 \beta +35\right) x_{ij} \left(\ai +\aj y_{ij}^2\right)+\left(3 \beta ^2+28 \beta +35\right) y_{ij} (\ai +\aj)\biggr] \\
  \mathcal{D}_{\text{SQ}1}^{\text{nG}1} = \,& \frac{2}{3} \left( 3 + \beta\right) \left( \frac{1}{\ai} + \frac{1}{\aj} \right)\\
  \mathcal{D}_{\text{NLB}}^{\text{nG}1} = \,&\mathcal{D}_{\text{SQ}1}^{\text{nG}1} \\
  \mathcal{D}_{\text{SQ2}}^{\text{nG}1} = \,& \frac{2 \beta}{15 \ai \aj \left(2 x_{ij}
  y_{ij}+y_{ij}^2+1\right)}  \biggr[\ai \left(\beta +2 \beta  x_{ij}^2+2 (3 \beta +5) x_{ij} y_{ij}+(3 \beta +5) y_{ij}^2+5\right)+ \nonumber \\  & + \aj \left(3 \beta +2 \beta 
   x_{ij}^2 y_{ij}^2+2 (3 \beta +5) x_{ij} y_{ij}+(\beta +5) y_{ij}^2+5\right)\biggr]\\
  \mathcal{D}_{\text{FOG}}^{\text{nG}1} = \,& \frac{\beta}{105 \ai \aj
  y_{ij}} \biggr[6 \beta ^2 x_{ij}^3 \left(\ai+\aj y_{ij}^2\right)+6 \beta  (4 \beta +7) x_{ij}^2 y_{ij} (\ai+\aj)+ \nonumber \\ \nonumber & \qquad \qquad + x_{ij} \biggr((\ai \left(9 \beta ^2+42 \beta +\left(15 \beta ^2+42 \beta +35\right) y_{ij}^2+35\right)+ \\ \nonumber & \qquad \qquad + \aj \left(15 \beta ^2+42 \beta +\left(9 \beta^2+42 \beta +35\right) y_{ij}^2+35\right)\biggr)+ \\ & \qquad \qquad 2 \left(3 \beta ^2+21 \beta +35\right) y_{ij} (\ai + \aj)\biggr] \\
   \mathcal{D}_{\text{nG}2}^{\text{nG}1} = \,&\mathcal{D}_{\text{SQ}1}^{\text{nG}1} \\
  \mathcal{D}_{\text{FoGnG}}^{\text{nG}1} = \,& \frac{\beta}{15 \ai^2 \aj^2 y_{ij}}  \biggr[4 \ai \aj \beta  x_{ij}^2 y_{ij}+ \nonumber \\& \qquad \qquad +(3 \beta +5) x_{ij} \left(\ai^2+\ai \aj
   \left(y_{ij}^2+1\right)+\aj^2 y_{ij}^2\right)+ \nonumber \\& \qquad \qquad + y_{ij} \left(\ai^2 (3 \beta +5)+2 \ai \aj (\beta +5)+\aj^2 (3 \beta +5)\right)\biggr] \\
  \mathcal{D}_{\text{SQ2}}^{\text{nG}1^2} = \,& \frac{2}{3} \frac{\beta}{\ai \aj}\\
  \mathcal{D}_{\text{FOG}}^{\text{nG}1^2} = \,& \frac{\beta}{15 \ai \aj y_{ij}}  \biggr[ 4 \beta x_{ij}^2 y_{ij}+(3 \beta +5) x_{ij} \left(y_{ij}^2+1\right)+2 (\beta +5) y_{ij}\biggr] \\
  \mathcal{D}_{\text{FoGnG}}^{\text{nG}1^2}  = \,& \frac{\beta}{3 \ai^2 \aj^2 y_{ij}}  \biggr[ x_{ij} \left(\ai+\aj y_{ij}^2\right)+y_{ij} (\ai+\aj)\biggr] \,,
\end{align}
where $x_{ij} = (\k_i \cdot \k_j)/{k_i k_j}$ and $y_{ij}=k_i/k_j$.

\section{Basic numbers for BOSS, eBOSS, DESI, Euclid }\label{app:data}
Here we present tables with the numbers describing the BOSS, DESI, Euclid \cite{Font-Ribera:2013rwa} and eBOSS \cite{Zhao:2015gua} surveys, which we use to forecast constraints on primordial non-Gaussianity in \cref{sec:fisher}. 

\begin{table}[hb!]
\centering
\caption{Basic numbers for BOSS LRGs. The shell volume $V$ is in units of $(\text{Gpc}/h)^3$, while the number density $N_{\text{LRG}}$ in $10^{-4}\,(h/\text{Mpc})^3$. The fiducial value for $b_{10}^{\text{LRG}}$ and the estimates of $\nu_{\text{LRG}}$ and the non-linear bias $b_{20}^{\text{LRG}}$ are also presented.}
\label{tab:BOSS} 
\begin{tabular}{cccccc}
\hline
\hline
$z$ & $	V $ & $ N_{\text{LRG}} $ & $ b_{10}^{\text{LRG}} $ & $ \nu_{\text{LRG}} $ & $ b_{20}^{\text{LRG}} $ \\
\hline
0.05 &	 0.03 &	 3.14 &	 1.74 &	 1.68 &	 -0.04 \\
0.15 &	 0.16 &	 3.06 &	 1.84 &	 1.74 &	  0.02 \\
0.25 &	 0.40 &	 3.12 &	 1.94 &	 1.81 &	  0.09 \\
0.35 &	 0.70 &	 3.17 &	 2.04 &	 1.88 &	  0.18 \\
0.45 &	 1.03 &	 3.21 &	 2.15 &	 1.95 &	  0.29 \\
0.55 &	 1.38 &	 3.25 &	 2.26 &	 2.01 &	  0.41 \\
0.65 &	 1.71 &  1.22 &	 2.37 &	 2.08 &	  0.55 \\
0.75 &	 2.03 &	 0.15 &	 2.49 &	 2.15 &	  0.70 \\
\hline
\hline
\end{tabular}
\end{table}

\begin{table}[hb!]
\centering
\caption{Basic numbers for eBOSS LRGs. The shell volume $V$ is in units of $(\text{Gpc}/h)^3$, while the number density $N_{\text{LRG}}$ in $10^{-4}\,(h/\text{Mpc})^3$. The fiducial value for $b_{10}^{\text{LRG}}$ and the estimates of $\nu_{\text{LRG}}$ and the non-linear bias $b_{20}^{\text{LRG}}$ are also presented.}
\label{tab:eBOSS_LRGs} 
\begin{tabular}{ccccccc}
\hline
\hline
$z$ & $	V $ & $ N_{\text{LRG}} $ & $ b_{10}^{\text{LRG}} $ & $ \nu_{\text{LRG}} $ & $ b_{20}^{\text{LRG}} $ \\
\hline
0.65 &	 1.20 & 0.810 &	 2.37 &	 2.08 &	  0.55 \\
0.75 &	 1.42 & 0.678 &	 2.49 &	 2.15 &	  0.70 \\
0.85 &	 1.63 & 0.350 &	 2.61 &	 2.21 &	  0.87 \\
0.95 &	 1.82 & 0.097 &	 2.73 &	 2.28 &	  1.06 \\
\hline
\hline
\end{tabular}
\end{table}

\begin{table}[!htb!]
\centering
\caption{Basic numbers for eBOSS QSOs. The shell volume $V$ is in units of $(\text{Gpc}/h)^3$, while the number density $N_{\text{QSO}}$ in $10^{-4}\,(h/\text{Mpc})^3$. The fiducial value for $b_{10}^{\text{QSO}}$ and the estimates of $\nu_{\text{QSO}}$ and the non-linear bias $b_{20}^{\text{QSO}}$ are also presented.}
\label{tab:eBOSS_QSO} 
\begin{tabular}{cccccc}
\hline
\hline
$z$ & $	V $ & $ N_{\text{QSO}} $ & $ b_{10}^{\text{QSO}} $ & $ \nu_{\text{QSO}} $ & $ b_{20}^{\text{QSO}} $ \\
\hline
0.65 &	 1.28 &	 0.119 &	 1.32 &	 1.33 &	 -0.22 \\
0.75 &	 1.52 &	 0.130 &	 1.42 &	 1.42 &	 -0.19 \\
0.85 &	 1.74 &	 0.154 &	 1.52 &	 1.51 &	 -0.16 \\
0.95 &	 1.95 &	 0.171 &	 1.63 &	 1.59 &	 -0.11 \\
1.05 &	 2.12 &	 0.163 &	 1.75 &	 1.68 &	 -0.04 \\
1.15 &	 2.28 &	 0.170 &	 1.87 &	 1.77 &	  0.04 \\
1.30 &	 4.96 &	 0.175 &	 2.06 &	 1.89 &	  0.21 \\
1.50 &	 5.36 &	 0.166 &	 2.34 &	 2.06 &	  0.51 \\
1.70 &	 5.65 &	 0.151 &	 2.64 &	 2.23 &	  0.93 \\
1.90 &	 5.84 &	 0.137 &	 2.97 &	 2.40 &	  1.48 \\
2.05 &	 2.96 &	 0.122 &	 3.23 &	 2.53 &	  1.99 \\
2.15 &	 2.98 &	 0.093 &	 3.41 &	 2.61 &	  2.39 \\	
\hline
\hline
\end{tabular}
\end{table}

\begin{table}[!htb!]
\centering
\caption{Basic numbers for eBOSS ELGs. The labels Fisher, LD, HD stand respectively for Fisher Discriminant, Low Density DECam and High Density DECam selected objects. The shell volume $V$ is in units of $(\text{Gpc}/h)^3$, while the expected number density $N_X$ based on target selection definition $X$ is in $10^{-4}\,(h/\text{Mpc})^3$. The fiducial value for $b_{10}^{\text{ELG}}$ and the estimates of $\nu_{\text{ELG}}$ and the non-linear bias $b_{20}^{\text{ELG}}$ are also presented.}
\label{tab:eBOSS_ELGs}
\begin{tabular}{cccccccccccccc}
\hline
\hline
$z$ & $	V $ & $ N_{\text{Fisher}} $ & $ N_{\text{LD}} $ & $ N_{\text{HD}} $ & $ b_{10}^{\text{ELG}} $ & $ \nu_{\text{ELG}} $ & $ b_{20}^{\text{ELG}}$ \\
\hline
0.65 &	 0.26 &	 1.41 &  0.183 	& 0.205	& 1.40 	& 1.40 &	 -0.20 \\	
0.75 &	 0.30 &	 2.17 &  1.91	& 2.07  & 1.46 	& 1.46 &	 -0.18 \\
0.85 &	 0.35 &	 1.65 &  2.67	& 3.03	& 1.53 	& 1.51 &	 -0.15 \\	
0.95 &	 0.39 &	 0.624 & 1.14	& 1.61	& 1.60 	& 1.57 &	 -0.12 \\	
1.05 &	 0.42 &	 0.218 & 0.373	& 0.568	& 1.68 	& 1.63 &	 -0.08 \\ 
1.15 &	 0.46 &	 0.081 & 0.159 	& 0.241	& 1.75 	& 1.68 &	 -0.04 \\
\hline
\hline
\end{tabular}
\end{table}

\begin{table}[!htb!]
\centering
\caption{Basic numbers for DESI. The shell volume $V$ is in units of $(\text{Gpc}/h)^3$, while the number density $N_X$ for the tracers $X$ (LRGs, ELGs, QSOs) is in $10^{-4}\,(h/\text{Mpc})^3$. The corresponding fiducial value for $b^X_{10}$ and the estimates of $\nu_X$ and the non-linear bias $b^X_{20}$ are also presented.}
\label{tab:DESI} 
\resizebox{\columnwidth}{!}{
\begin{tabular}{cccccccccccccc}
\hline
\hline
$z$ & $	V $ & $ N_{\text{ELG}} $ & $ b_{10}^{\text{ELG}} $ & $ \nu_{\text{ELG}} $ & $ b_{20}^{\text{ELG}} $ & $ N_{\text{LRG}} $ & $ b_{10}^{\text{LRG}} $ & $ \nu_{\text{LRG}} $ & $ b_{20}^{\text{LRG}} $ & $ N_{\text{QSO}} $ & $ b_{10}^{\text{QSO}} $ & $ \nu_{\text{QSO}} $ & $ b_{20}^{\text{QSO}} $\\
\hline
0.15 &	 0.23 &	  23.0 &  0.91 	&	 0.85 &	 -0.21 	&	3.06 	& 1.84 	& 	1.74 	&  0.02	   &	 0.489 	& 1.30  	& 1.31 	&	-0.22 	  \\
0.25 &	 0.56 &   8.65 &  0.96 	& 	 0.92 &	 -0.22	&	3.12 	& 1.94 	& 	1.81 	&  0.09    &	 0.574 	& 1.37 	        & 1.37 	&	-0.21 	  \\
0.35 &	 0.98 &	  4.15 &  1.01 	& 	 0.99 &	 -0.23 	&	3.17 	& 2.04 	& 	1.88 	&  0.18    &	 0.442 	& 1.44 	       	& 1.44 	& 	-0.19 	  \\
0.45 &	 1.45 &	  2.76 &  1.06 	& 	 1.06 &	 -0.23	&	3.21 	& 2.15 	&	1.95 	&  0.29    &	 0.300 	& 1.52 	 	& 1.50 	& 	-0.16 	  \\
0.55 &	 1.93 &	  3.13 &  1.12 	& 	 1.12 &	 -0.23	&	3.26 	& 2.26 	&	2.01 	&  0.41    &	 0.233 	& 1.59 	 	& 1.56 	& 	-0.13 	  \\
0.65 &	 2.40 &	  4.22 &  1.17 	& 	 1.18 &	 -0.23 	&	3.29 	& 2.37 	&	2.08 	&  0.55    &	 0.199 	& 1.67 	 	& 1.62 	& 	-0.08     \\
0.75 &	 2.84 &	  5.48 &  1.23  &	 1.24 &	 -0.23	&	3.32 	& 2.49 	&	2.15 	&  0.70    &	 0.182 	& 1.76 	 	& 1.68 	& 	-0.04 	  \\
0.85 &	 3.26 &	  5.73 &  1.29 	&	 1.30 &	 -0.22 	&	2.03 	& 2.61 	&	2.21 	&  0.87    &	 0.189 	& 1.84 	 	& 1.74 	& 	 0.02 	  \\
0.95 &	 3.63 &	  5.40 &  1.35 	&	 1.35 &	 -0.21	&	0.35 	& 2.73 	&	2.28 	&  1.06    &	 0.193 	& 1.93 	 	& 1.80 	& 	 0.09 	  \\
1.05 &   3.97 &	  5.19 &  1.41 	&	 1.41 &	 -0.20 	&	0.04 	& 2.85 	&	2.34 	&  1.26    &     0.198 	& 2.01 	 	& 1.86 	& 	 0.16 	  \\
1.15 &	 4.26 &	  4.87 &  1.47 	&	 1.46 &	 -0.18  &	0 	& 0 	&	0 	&  0       &	 0.204 	& 2.10 	 	& 1.92 	& 	 0.24 	  \\
1.25 &	 4.52 &	  4.40 &  1.53 	&	 1.51 &	 -0.15	&	0 	& 0 	&	0 	&  0       &	 0.214 	& 2.19 	 	& 1.97 	& 	 0.33 	  \\
1.35 &	 4.74 &	  3.31 &  1.59 	&	 1.56 &	 -0.13	&	0 	& 0 	&	0 	&  0       &	 0.222 	& 2.27 	 	& 2.02 	& 	 0.43     \\
1.45 & 	 4.93 &	  2.20 &  1.65 	&	 1.61 &	 -0.10	&	0 	& 0 	&	0 	&  0       &	 0.230 	& 2.36 	 	& 2.08 	& 	 0.54     \\
1.55 & 	 5.09 &	  1.27 &  1.72 	&	 1.66 &	 -0.06	&	0 	& 0 	&	0 	&  0       &	 0.228 	& 2.45 	 	& 2.13 	& 	 0.65     \\
1.65 & 	 5.22 &	 0.480 &  1.78 	&	 1.70 &	 -0.02	&	0 	& 0 	&	0 	&  0       &	 0.215 	& 2.54 	 	& 2.18 	& 	 0.78     \\
1.75 & 	 5.33 &	 0.129 &  1.84 	&	 1.75 &	  0.02 	&	0 	& 0 	&	0 	&  0       &	 0.202 	& 2.63 	 	& 2.23 	& 	 0.91     \\
1.85 & 	 5.41 &	   0   &  0	&	 0    &	  0	&	0 	& 0 	&	0 	&  0       &	 0.191 	& 2.72 	 	& 2.28 	& 	 1.05     \\
\hline
\hline
\end{tabular}}
\end{table}

\begin{table}[!htb!]
\centering
\caption{Basic numbers for Euclid. The shell volume $V$ is in units of $(\text{Gpc}/h)^3$, while the number density $N$ in $10^{-4}\,(h/\text{Mpc})^3$. The fiducial value for $b_{10}$ and the estimates of $\nu$ and the non-linear bias $b_{20}$ are also presented.}
\label{tab:Euclid} 
\begin{tabular}{cccccc}
\hline
\hline
$z$ & $	V $ & $ N $ & $ b_{10} $ & $ \nu $ & $ b_{20} $ \\
\hline
0.65 &	 2.57 &	 6.42  & 1.06 &	 1.06 &	 -0.23 \\
0.75 &	 3.05 &	 14.5  & 1.11 &	 1.12 &	 -0.23 \\
0.85 &	 3.49 &	 16.3  & 1.17 &	 1.18 &	 -0.23 \\
0.95 &	 3.89 &	 15.0  & 1.22 &	 1.23 &	 -0.23 \\
1.05 &	 4.25 &	 13.3  & 1.27 &	 1.29 &	 -0.22 \\ 
1.15 &	 4.57 &	 11.6  & 1.33 &	 1.34 &	 -0.21 \\
1.25 &	 4.84 &	 10.1  & 1.38 &	 1.39 &	 -0.20 \\
1.35 &	 5.08 &	 8.42  & 1.44 &	 1.44 &	 -0.19 \\ 
1.45 &	 5.28 &	 6.68  & 1.50 &	 1.48 &	 -0.17 \\
1.55 &	 5.45 &	 5.09  & 1.55 &	 1.53 &	 -0.14 \\
1.65 &	 5.59 &	 3.69  & 1.61 &	 1.58 &	 -0.12 \\ 
1.75 &	 5.71 &	 2.56  & 1.67 &	 1.62 &	 -0.09 \\
1.85 &	 5.80 &	 1.68  & 1.73 &	 1.66 &	 -0.05 \\
1.95 &	 5.87 &	 1.02  & 1.78 &	 1.70 &	 -0.02 \\ 
2.05 &	 5.93 &	 0.380 & 1.84 &	 1.74 &	  0.02 \\
\hline
\hline
\end{tabular}
\end{table}

\end{document}